\documentclass[structabstract]{aa}  

\usepackage{color}
\usepackage{natbib}                     
\usepackage{graphicx}                   
\usepackage[varg]{txfonts}              
\usepackage[english]{babel}             
\usepackage{nameref}                    
\usepackage{paralist}                   
\usepackage{multirow}

\newcommand{\teff}  {T$_\mathrm{eff}$}
\newcommand{\logg}  {$\log g$}
\newcommand{\kms}{\hbox{${\rm km}\:{\rm s}^{-1}$}}

\begin{document}

  \title{Searching for Li-rich giants in a sample of 12 open clusters\thanks{
Based on observations collected at the La Silla Observatory, ESO
(Chile), with HARPS/3.6m (runs ID 075.C-0140, 076.C-0429, 077.C-0088, and 078.C-0133) and with UVES/VLT 
at the Cerro Paranal Observatory (run 079.C-0131)}}
\subtitle{Li enhancement in two stars with substellar companions}

  \author{E. Delgado Mena\inst{1}
          \and M. Tsantaki\inst{1}
          \and S. G. Sousa\inst{1}
          \and M. Kunitomo\inst{2}
          \and V. Adibekyan\inst{1}
          \and P. Zaworska\inst{3}
          \and N. C. Santos\inst{1,3}
          \and G. Israelian\inst{4,5}
          \and C. Lovis\inst{6}
         }

  \institute{
          Instituto de Astrof\'isica e Ci\^encias do Espa\c{c}o, Universidade do Porto, CAUP, Rua das Estrelas, 4150-762 Porto, Portugal
          \email{Elisa.Delgado@astro.up.pt}
          \and
          Department of Physics, Nagoya University, Furo-cho, Chikusa-ku, Nagoya, Aichi 464-8602, Japan
          \and
          Departamento de F\'isica e Astronomia, Faculdade de Ci\^encias, Universidade do Porto, Rua do Campo Alegre, 4169-007 Porto, Portugal
          \and
          Instituto de Astrof\'{\i}sica de Canarias, 38200 La Laguna, Tenerife, Spain
          \and 
          Departamento de Astrof{\'\i}sica, Universidad de La Laguna, 38206 La Laguna, Tenerife, Spain
          \and
          Observatoire de Gen\'eve, Université de Gen\'eve, 51 ch. des Maillettes, 1290 Sauverny, Switzerland
}

  \date{Received date / Accepted date }
 
  \abstract
  {}
  {The aim of this work is to search for Li-rich giants in a sample of clusters where planets have been searched, thus we can study the planet engulfment scenario to explain Li replenishment using a proper comparison sample of stars without detected giant planets.}
  {We derived Li abundances for a sample of 67 red giant stars in 12 different open clusters using standard spectral synthesis techniques and high-resolution spectra (from HARPS and UVES). We also determined masses, ages, and radius from PARSEC stellar isochrones to constrain the evolutionary stage of these stars.}
  {We found three stars in different clusters with clearly enhanced Li abundances compared to other stars within the cluster. Interestingly, the only two stars with a detected substellar companion in our sample belong to that group. One of the planet hosts, NGC2423No3, might lie close to the luminosity bump on the HR diagram, a phase where Li production by the Cameron-Fowler process is supported by extra-mixing to bring fresh Li up to the surface. On the other hand, NGC4349No127 is a more massive and more evolved giant that does not seem to be in the evolutionary phase where other Li-rich stars are found. We discuss the possibility that the Li enhancement of this star is triggered by the engulfment of a planet, considering that close-in planets hardly survive the RGB tip and the early AGB phases.}
  {}
  \keywords{stars:~abundances -- stars:~planetary systems -- stars:~rotation -- stars:~evolution -- planets and satellites:~physical evolution}

  \maketitle
%
\section{Introduction}                                  \label{sec:Introduction}
The abundances of light elements, for example Li, can suffer dramatic changes along the evolution of a star, the opposite to what is expected for most of the heavier elements. The abundances of Li for main-sequence (MS) stars are mainly governed by the stellar effective temperature (\teff) and can undergo a smooth decrease with age after the first 2 Gyr of their lives \citep{sestito05}. The stars cooler than $\sim$\,5600K show very depleted Li abundances due to their thicker convective envelopes. Contrarily, for stars hotter than the Sun, the Li abundances increase with \teff\ until they reach a temperature of around 6400K \citep[e.g.][]{delgado15} where they undergo a sudden decrease to form the so-called Li-dip \citep{boesgaard86}. However, the Li abundances of the stars at the hot side of the Li-dip hardly vary during the MS. Once the stars leave the MS they experience the first dredge-up (FDU) when the convective envelope deepens and then recedes. At this moment the Li that had been conserved in the outer convective zone is mixed with the hotter interior triggering a drop in Li abundances typically below A(Li)\footnote{${\rm A(Li)}=\log[N({\rm Li})/N({\rm H})]+12$}\,=\,1.5\,dex \citep[][hereafter CB00]{charbonnel00}. Thus, the stars in the red giant branch (RGB) are expected to show low Li abundances, though a number of Li-rich giants have been observed in clusters and in the field \citep[e.g.][]{kumar11,carlberg12,anthony-twarog13,d'orazi15}, sometimes exceeding the meteoritic values.\\

Several scenarios have been proposed to explain the unexpected high Li abundances of these objects. On the one hand, an external contamination by accretion of planetary material \citep{alexander67,siess99} or by accretion of interstellar gas enriched by supernova explosions \citep{woosley95} might increase the Li content in the photosphere. On the other hand, the internal production of Li through the Cameron-Fowler mechanism (CFM) \citep{cameron-fowler} in the $^3$He rich shells of stars more massive than 1.5\,M$_\odot$ can lead to an increase in abundance. For stars with masses of 4-6 M$_\odot$ on the asymptotic giant branch (AGB), the H-burning shell where the reaction $^3$He($\alpha,\gamma$)$^7$Be takes place, is close to the base of the convective zone. Then, the recently created $^7$Be can quickly reach the cooler outward regions to decay to $^7$Li through Hot Bottom Burning \citep{sackmann92}. However, for less massive giants a non-standard mixing process is needed to connect the H-burning shell to the convective envelope in order to bring the fresh Li up to the photosphere \citep{sackmann99}. Many of the low-mass ($<$\,2-2.3\,M$_\odot$) RGB Li-rich stars are observed close to the luminosity bump where they show a brief phase of decreasing luminosity when ascending through the RGB (CB00). At this evolutionary phase, the outward hydrogen burning shell reaches the mean molecular weight discontinuity created by the FDU. The decrease in the $^{12}$C/$^{13}$C that begins to take place at this phase is probably caused by some extra-mixing processes, which in turn support the CFM to enhance Li in these stars \citep[CB00,][]{eggleton08}. On the other hand, \citet{kumar11} claim that the highest concentration of Li-rich stars happens at the red clump rather than the luminosity bump. Then, either the fresh Li produced during the bump can survive the RGB tip or the CFM can operate at the He-flash. This last suggestion is supported by the asteroseismic analysis of a red giant burning He in its core \citep{silva-aguirre14} and by Li-rich red clump stars discovered by \citet{monaco14} and \citet{strassmeier15}. Finally, other recent works report Li-rich stars all along the RGB challenging the idea that Li production is only possible at certain evolutionary phases \citep{gonzalez_lirich09,monaco11,lebzelter12}. \\

The disadvantage of many of the previously mentioned studies is the difficulty in determining the evolutionary stage of field stars due to the associated errors on derived ages and masses unless asteroseismic observations are carried out. Moreover, many of the past Li abundance compilations were clearly not homogeneous. Therefore, the aim of this work is to search for Li-rich stars in a sample of homogenously analysed red giants belonging to clusters. By studying stars in clusters we can significantly constrain their evolutionary state and thus the mechanisms associated with the enrichment of Li. Moreover, the stars analysed in our sample have been observed during several periods in order to detect planets around them, hence we can explore the possible connection of planet engulfment with Li-rich giants. We organize our paper as follows. In Sects. \ref{sec:observations} and \ref{sec:abundances} we present the data and explain how they were analysed and in Sect. \ref{sec:results} we discuss the results. The conclusions are presented in Sect \ref{sec:conclusions}.

\begin{figure}
  \centering
  \includegraphics[width=1.0\linewidth]{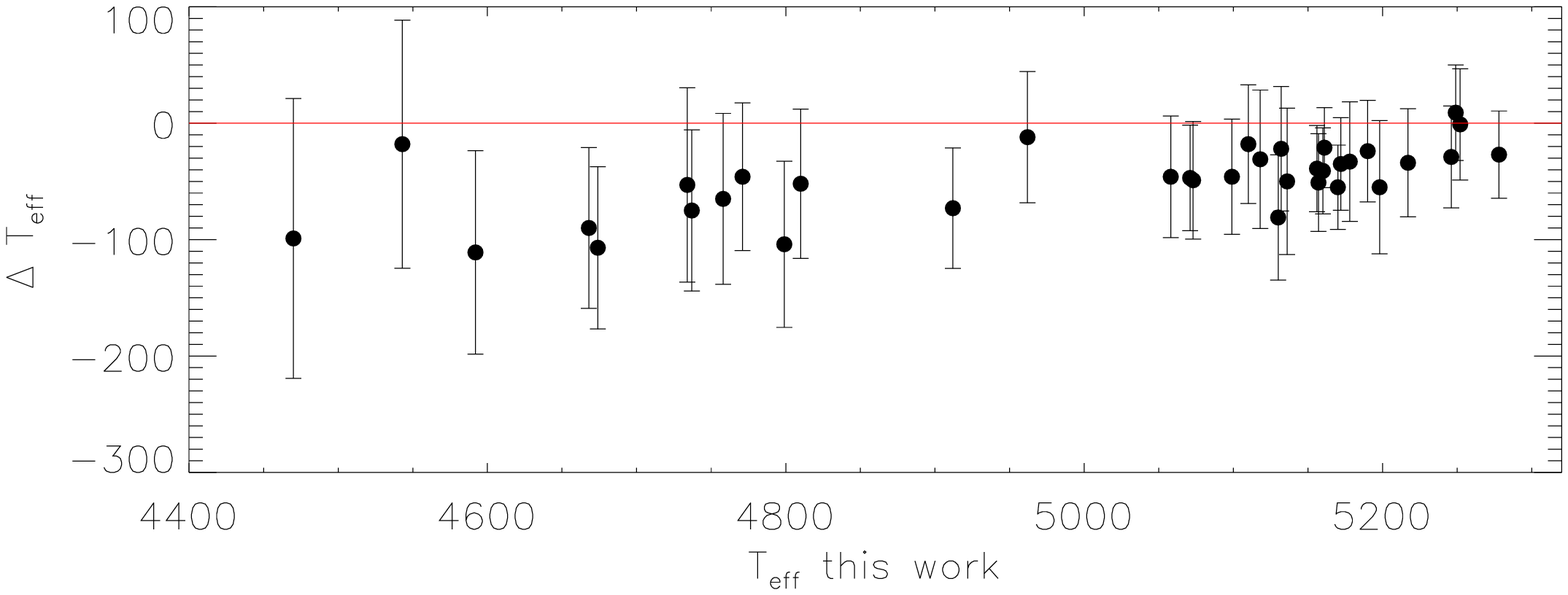}
  \includegraphics[width=1.0\linewidth]{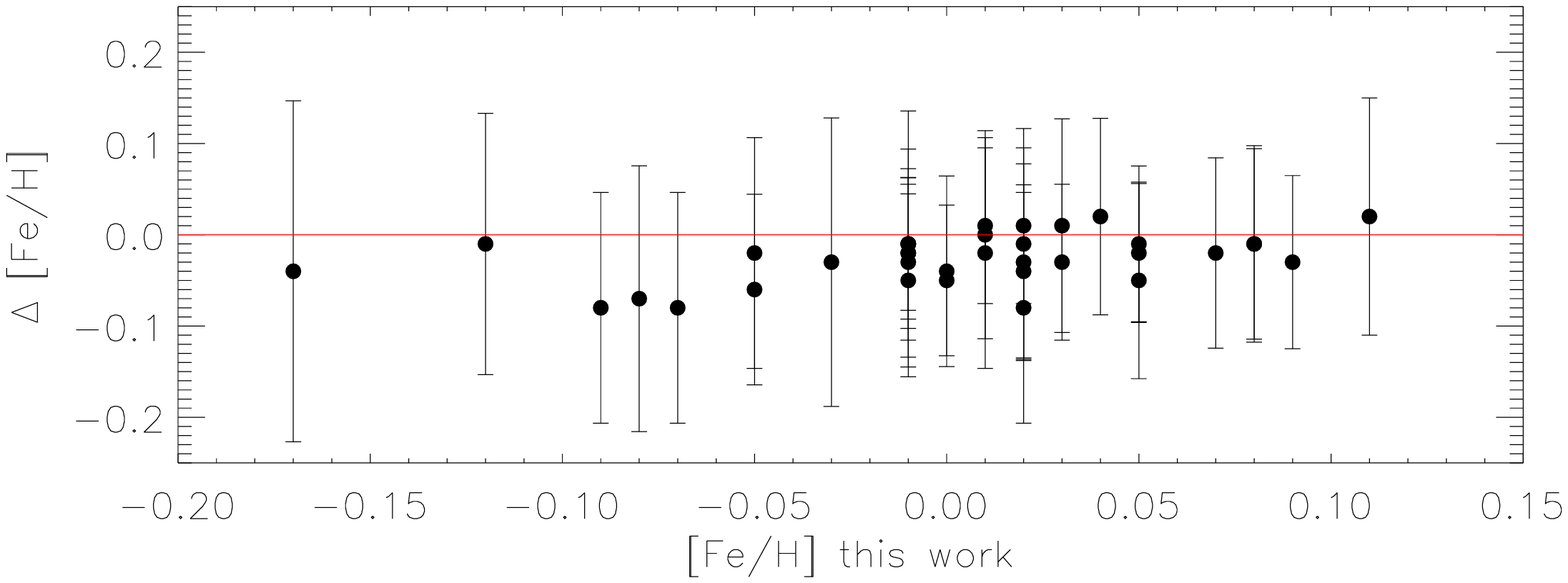}
  \includegraphics[width=1.0\linewidth]{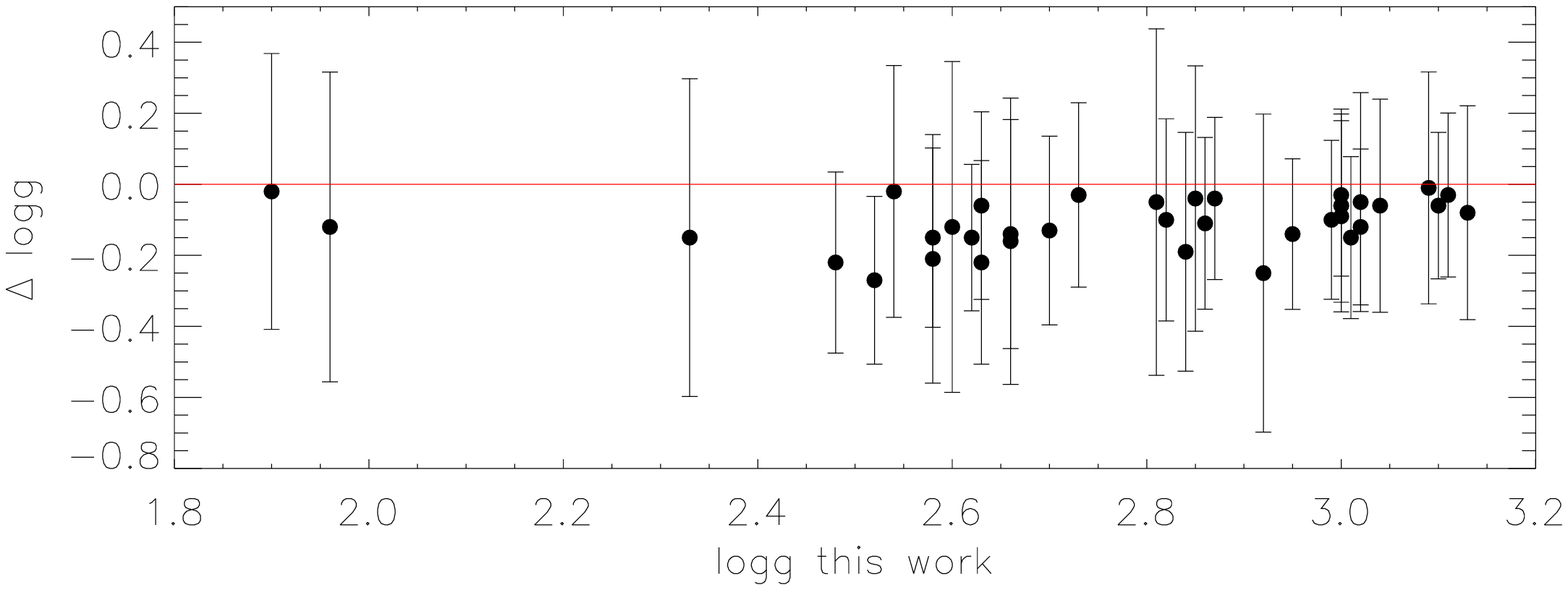}
  \caption{\footnotesize{Comparison of the parameters derived in this work and the results of \cite{santos09}: 
temperature (top panel), metallicity (middle panel), and surface gravity (bottom panel).}}
  \label{comparison_nuno}
  \end{figure}

\begin{figure}
  \centering
  \includegraphics[width=1.0\linewidth]{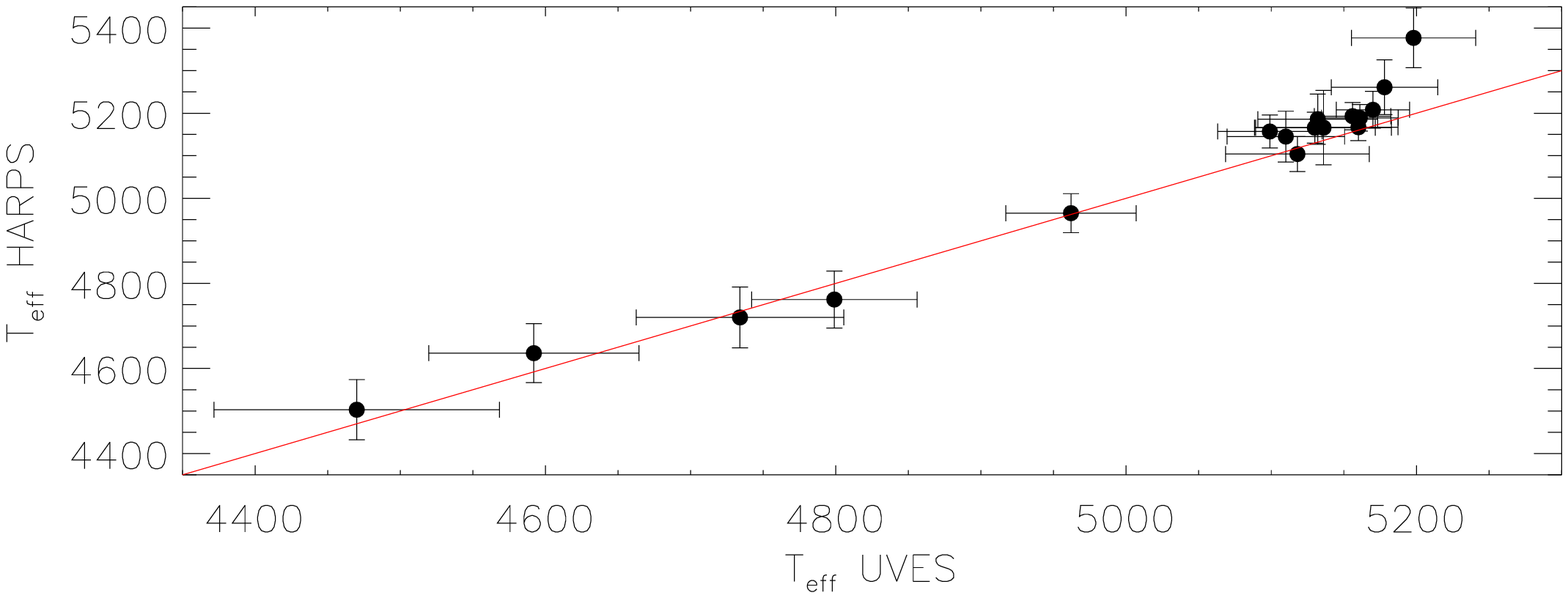}
  \includegraphics[width=1.0\linewidth]{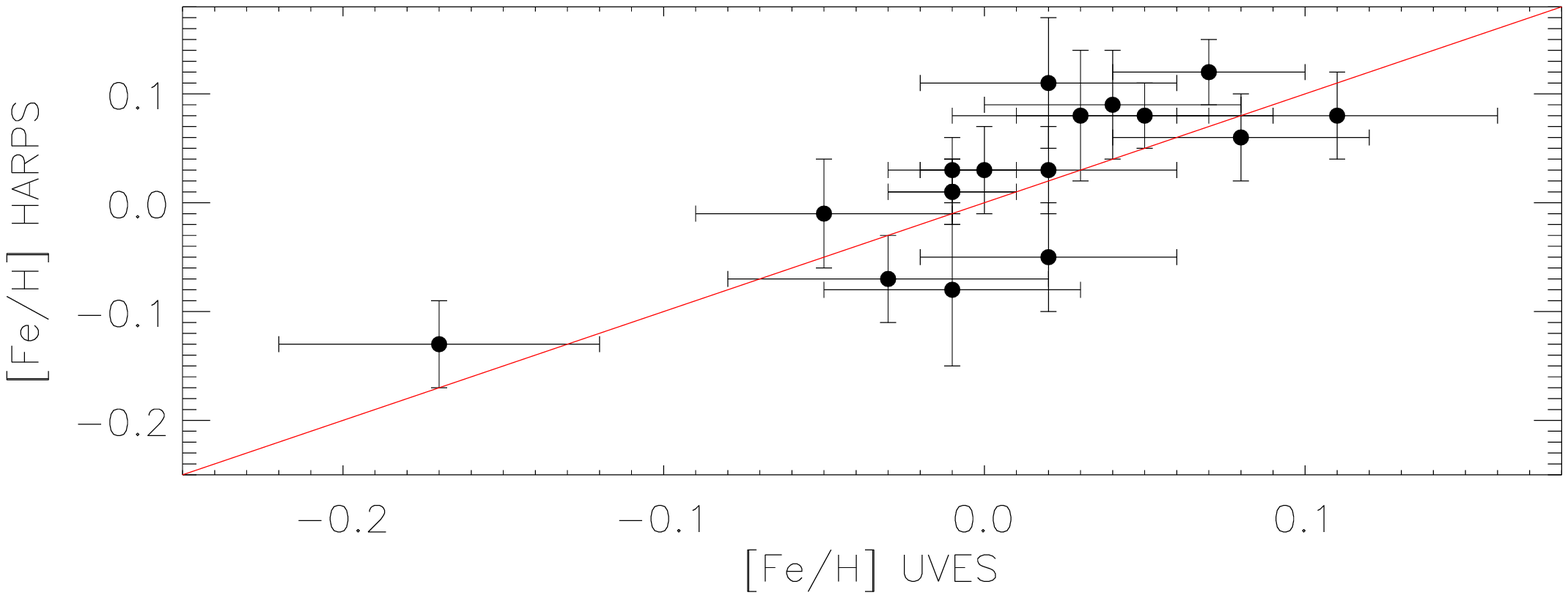}
  \caption{\footnotesize{Comparison of the parameters derived using HARPS or UVES spectra: 
temperature (top panel) and metallicity (middle panel)}}
  \label{comparison_resolution}
  \end{figure}

\section{Observations and stellar parameters} \label{sec:observations}

The baseline sample used in this work is taken from \citet{lovis07}, hereafter LM07. It is composed of 46 red giants in 12 different open clusters (see \citealt{lovis07} Table 1 of LM07) observed with the HARPS spectrograph at the ESO 3.6m telescope (R\,=\,115000). We included 21 additional giant stars from the same clusters observed by \citet{santos09} using the UVES/VLT spectrograph (R\,=\,50000).\\

The stellar parameters, \teff, surface gravity (\logg), metallicity ([Fe/H]), and 
microturbulence ($\xi_{t}$), were determined in a homogeneous way based on the equivalent width (EW) 
measurements of \ion{Fe}{i} and \ion{Fe}{ii} lines and by imposing iron excitation and ionization equilibrium. Although \citet{santos09} provide parameters for part of our sample, we prefer to re-derive them to ensure homogeneity using a specific line list for cool stars presented in \citet{tsantaki13}. This line list was specially selected to eliminate lines that suffer from blending effects which are strongly present for cooler stars ($<$5200\,K) and have a significant effect mainly on the determination of temperature. The obtained parameters are listed in Table \ref{parameters}. \\

The EW were calculated automatically using the code ARES2\footnote{http://www.astro.up.pt/$\sim$sousasag/ares} 
\citep{sousa_ares2} and the abundance analysis was performed in local thermodynamic equilibrium (LTE) using a grid of model atmospheres \citep[ATLAS9,][]{kurucz} and the radiative transfer code MOOG2010\footnote{For the latest version visit: http://verdi.as.utexas.edu/moog.html} \citep{sneden}. A comparison of the new parameters with the previous results of \citet{santos09} is shown in Fig.~\ref{comparison_nuno}. The overestimation of the effective temperature of the previous results is corrected with our line list. We also checked the consistency of our results considering the different resolution of our two sets of data. In Fig.~\ref{comparison_resolution}, we compare the parameters obtained for 17 stars observed with both HARPS and UVES (R\,=\,50000). The mean difference in \teff\ is 35\,$\pm$\,47\,K, (HARPS-UVES), a value within the errors. Only for one star do we obtain a difference above 100K. For [Fe/H] the differences are also small; the mean value is 0.014\,$\pm$\,0.045\,dex. A similar comparison was done by \citet{santos12} who derived parameters from UVES spectra with R\,=\,100000 and R\,=\,50000. They found similar metallicities, though the values tend to be a bit higher for higher resolution, as is true in our sample.\\ 

\begin{figure}
  \centering
  \includegraphics[width=1.0\linewidth]{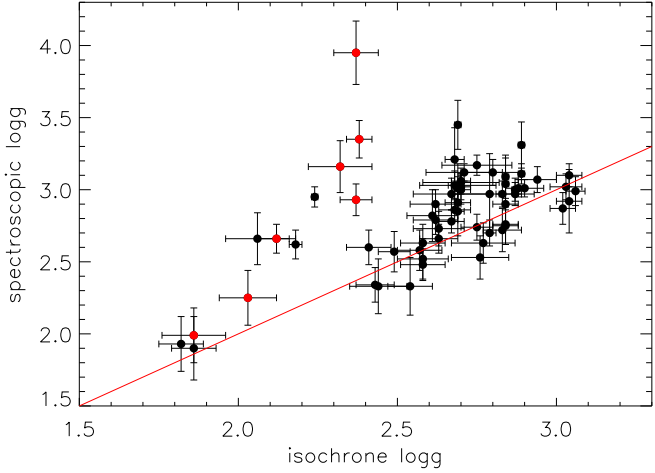}
  \caption{Spectroscopic \logg\ vs. \logg\ obtained from PARSEC isochrones. The stars from NGC4349 are overplotted with red symbols.}
  \label{logg}
  \end{figure}

Additionally, we derived the masses, radii, and ages from PARSEC v1.3 isochrones\footnote{http://stev.oapd.inaf.it/cgi-bin/param} \citep{bressan12} using the distances and $V$ magnitudes from WEBDA database\footnote{http://www.univie.ac.at/webda/}, 
and spectroscopic \teff\ and [Fe/H]. In order to get homogeneous and realistic ages (considering the youth of most of our clusters) for all the stars within a cluster, we had to constrain the maximum age with the Bayesian priors available in the PARSEC v1.3 webpage. As maximum values we used the ages given in Table 1 of LM07 plus the corresponding error. The comparison between the gravities from isochrones and the spectroscopic values shows a relatively good agreement where spectroscopic values are higher by 0.21\,$\pm$\,0.30\,dex (see Fig.\,\ref{logg}). However, for the cluster NGC4349 we found very high spectroscopic \logg\ for some of the stars, probably owing to their higher rotation rates. Therefore, we decided to use the \logg\ from isochrones for all the clusters in the rest of the analysis. Nevertheless, we note that Li abundances are only mildly sensitive to this difference in \logg. These parameters are shown in Table\,\ref{padova}.

\section{Determination of Li abundances} \label{sec:abundances}

\begin{figure}
\centering
\includegraphics[width=7.0cm,angle=270]{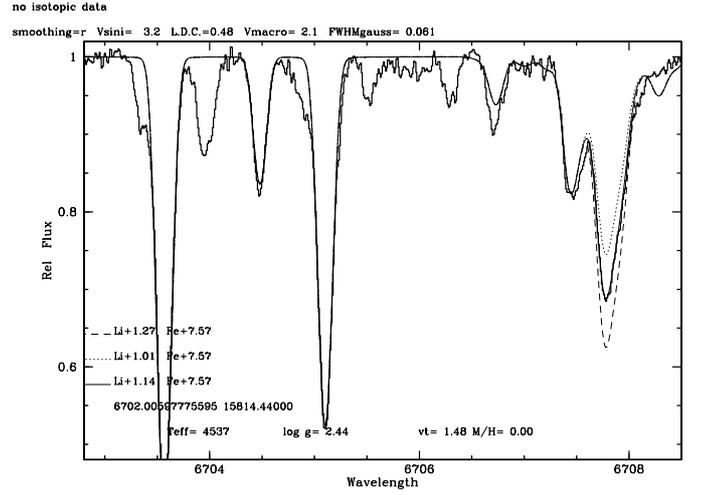}
\caption{Spectral synthesis around the Li line for IC4651No9791. The synthesis considering the errors in Li abundances are overplotted with dashed and dotted lines.} 
\label{IC4651_syn}
\end{figure}

We derived LTE lithium abundances by performing spectral synthesis with the code MOOG2010 \citep{sneden}, a grid of ATLAS9 atmospheres \citep{kurucz}, and the line list from \cite{ghezzi_li6} in the same way as derived by \cite{delgado14}. The obtained abundances are listed in Table\,\ref{parameters}. The Li synthesis for the five stars with enhanced Li are shown in Fig. \ref{IC4651_syn} and Appendix C.
The errors were computed by considering the variation in Li abundance due to the uncertainty in the continuum position and the sensitivity to the uncertainty in stellar parameters. To derive the second contribution we chose a set of stars with different \teff\ and detectable Li abundances and created new models with the original parameters excluding the one we wanted to check (which is varied with the corresponding error). Lithium abundances are barely sensitive to the changes in [Fe/H] ($\leq$\,0.01\.dex) and not sensitive to the changes in \logg\ and microturbulence. However, Li abundances change by 0.07-0.13\,dex for \teff\ uncertainties in the range 45-80\,K as shown in Table \ref{errors}. To derive the final error we added quadratically the error due to the continuum and the error due to \teff\ uncertainty for each star, taking as a reference stars with similar \teff\ and error from Table \ref{errors}.
Lithium abundances are known to be affected by NLTE (non-local thermodynamic equilibrium) corrections, thus we derived the corrected abundances using the tabulations from \citet{lind09}. Since all our stars are cool and have A(Li)\,$<$\,2\,dex, all the corrections are positive between 0.07 dex and 0.32 dex and produce higher abundances for the coolest stars.

\begin{center}
\begin{table}[h!]
\caption{Li abundance sensitivities to variations in \teff.}
\centering
\begin{tabular}{lcc}
\hline
\noalign{\smallskip} 
Star & $T_\mathrm{eff}$ &  $\Delta({\rm A(Li)})$  \\
\noalign{\smallskip} 
     & (K)  &  dex  \\
\noalign{\smallskip} 
\hline
\hline
\noalign{\smallskip} 
 NGC4349No127 &   4503 $\pm$ 70     &   0.13 \\
 IC4651No9791 &   4537 $\pm$ 72     &   0.12 \\
   NGC2423No3 &   4592 $\pm$ 72     &   0.10 \\
  NGC3680No13 &   4674 $\pm$ 56     &   0.07 \\
  NGC3680No26 &   4668 $\pm$ 47     &   0.07 \\
 IC4651No8540 &   4868 $\pm$ 72     &   0.10 \\
  IC2714No220 &   4980 $\pm$ 64     &   0.08 \\
 NGC2539No346 &   5104 $\pm$ 41     &   0.04 \\  
  IC2714No126 &   5211 $\pm$ 77     &   0.09 \\
   IC2714No87 &   5377 $\pm$ 70     &   0.10 \\
\hline
\noalign{\medskip} %
\end{tabular}
\end{table}
\label{errors}
\end{center}

\begin{figure}
\centering
\includegraphics[width=1.0\linewidth]{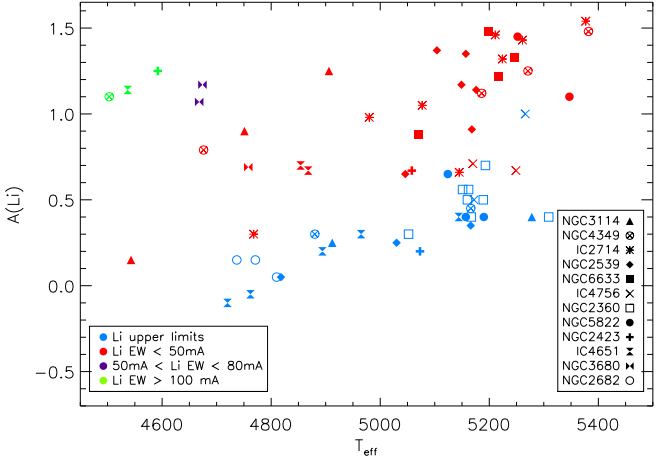}
\caption{LTE Li vs. \teff. Each cluster is represented by a different symbol, while the colours denote the strength of the Li line. The green symbols correspond to NGC4349No127 (circle with plus), IC4651No9791 (hourglass), and NGC2423No3 (diamond). The two purple symbols correspond to NGC3680No13/No26. } 
\label{Liteff}
\end{figure}

\section{Results and discussion} \label{sec:results}

\begin{figure}
\centering
\includegraphics[width=1.0\linewidth]{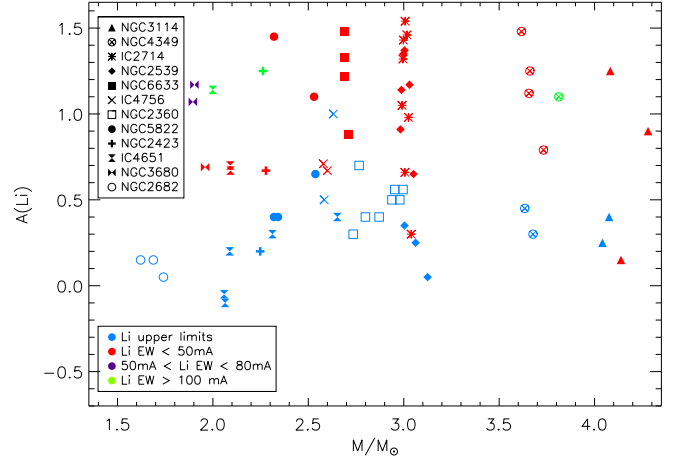}
\caption{LTE Li vs. stellar mass. Symbols as in Fig.\,\ref{Liteff}. } 
\label{Limass}
\end{figure}

In Fig.\,\ref{Liteff}, we show the derived LTE abundances of Li as a function of \teff. Lithium abundances increase with \teff, which might reflect the Li behaviour on the MS (hotter stars have thinner convective envelopes). We note that using NLTE abundances would produce the same trend, although the slope would be slightly less steep. A similar behaviour is found in \citet{gonzalez_lirich09} and \citet{monaco11}. However, to disentangle the possible causes for Li depletion we have to compare stars in the same evolutionary stage (see discussion below). We also divided the stars into four different groups depending on the strength of the Li absorption line. Assuming the standard definition of a Li-rich giant as a star with A(Li)\,$>$\,1.5\,dex, we find very few such stars in our sample. The most abundant stars belong to the hottest group of stars ($>$5000K), which in turn were the hottest stars during the MS and probably retained a higher amount of Li in their photospheres. Although the standard models do not predict Li depletion during the MS for stars with masses larger than 1.4 M$_\odot$, it seems that some depletion took place in the past since the abundances for some of our stars are much lower than predicted for stars arriving at the base of the RGB. This is more obvious in Fig.\,\ref{Limass} where a wide spread of abundances is observed for stars of a similar mass. However, for this subgroup of hotter stars, we cannot determine whether the spread in Li abundances originated on the MS or is due to a different degree of Li dilution during the FDU.

\subsection{Li-depleted stars}

\begin{figure}
\centering
\includegraphics[width=1.0\linewidth]{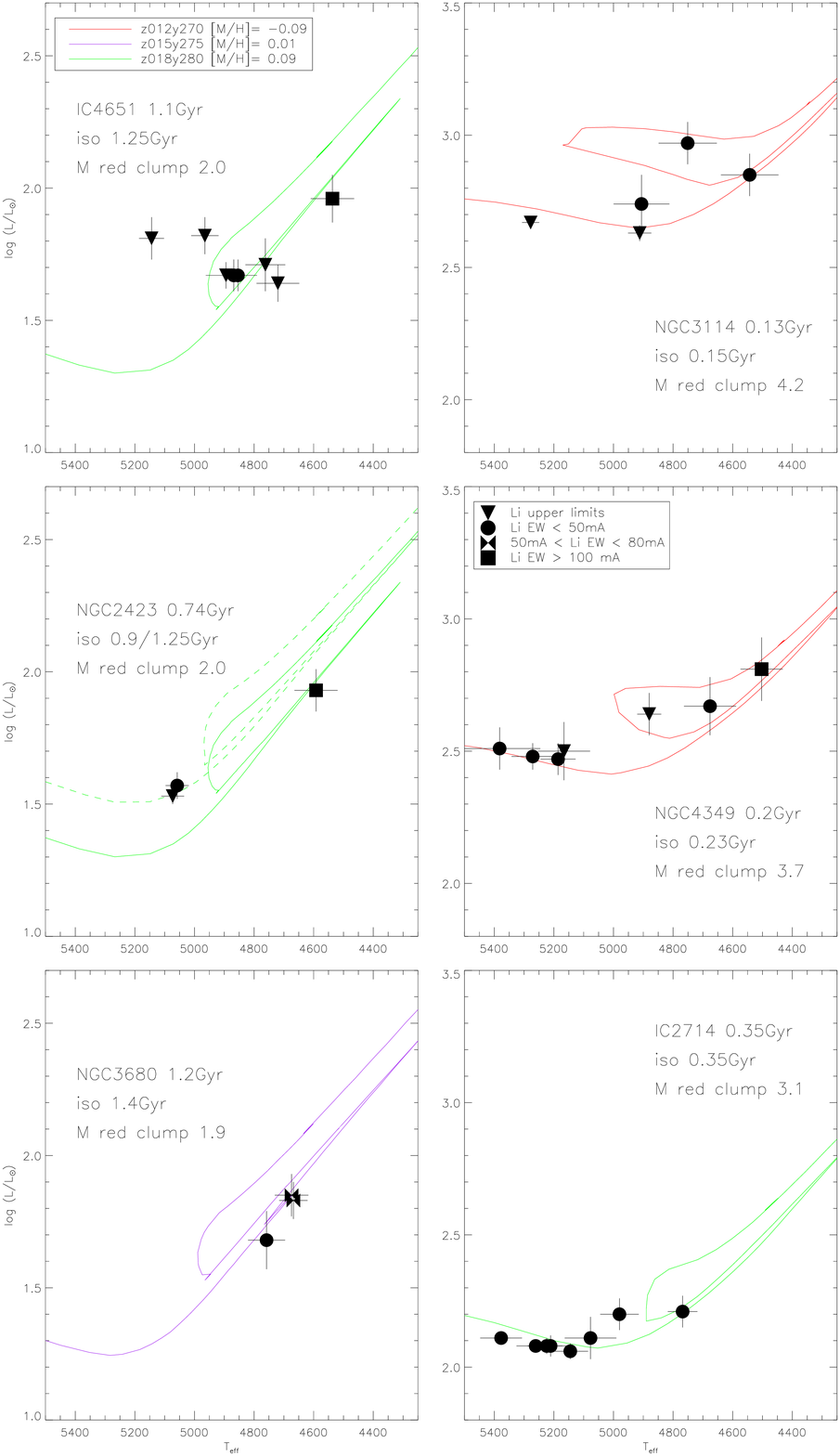}
\caption{HR diagram for half of the sample. The four different symbols correspond to the four different colours in Figs.\,\ref{Liteff} and \ref{Limass}. The Li-rich stars are depicted with filled squares and bowtie symbols. Colours denote different metallicities. In each panel the age of the cluster (given by WEBDA database) together with the ages of the isochrones that most closely matches our stars are depicted.} 
\label{HR1}
\end{figure}

\begin{figure}
\centering
\includegraphics[width=1.0\linewidth]{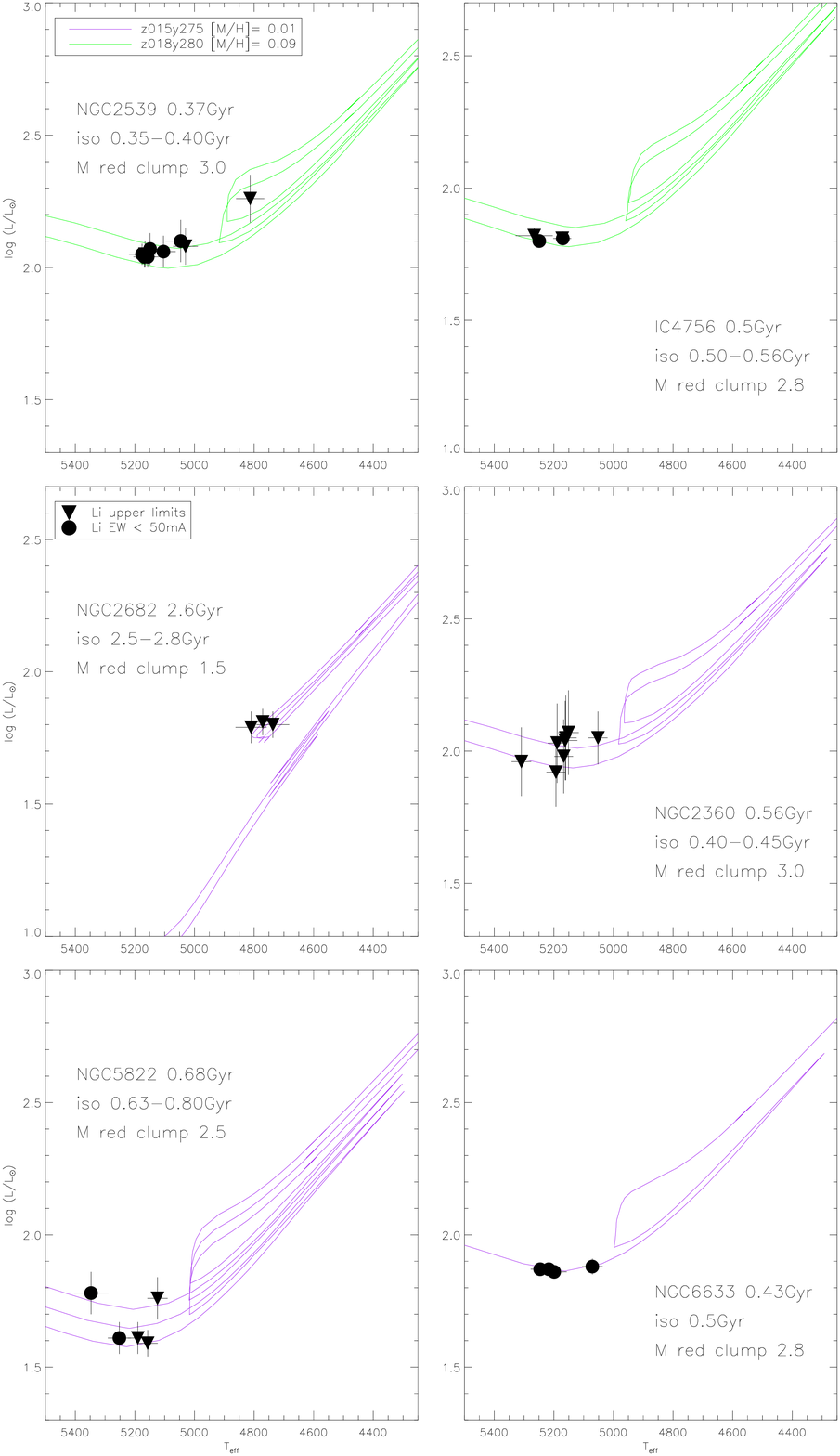}
\caption{Same as Fig. \ref{HR1} for the rest of the clusters} 
\label{HR2}
\end{figure}

In Figs.\,\ref{HR1} and \ref{HR2}, we show the Hertzsprung-Russell (HR) diagram for our clusters. For each cluster we show the PARSEC isochrones \citep{bressan12} that most closely match our stars considering the average metallicity of each cluster (using our [Fe/H] determination) and the age estimates from WEBDA database (see also Table 1 in LM07). Most of our stars are close to the first ascent on the RGB or at more evolved stages, so we expect them to have already experienced important Li dilution due to the FDU. However, the hottest objects in each cluster (i.e. least evolved), have probably experienced less dilution since they are at an earlier phase of the FDU and thus might be incorrectly catalogued as Li-rich or Li-enhanced giants, as also shown by CB00. This is probably the case of the three hottest stars with A(Li)\,$>$\,1.1 in NGC4349 that lie at the base of the RGB (see middle right panel of Fig.\,\ref{HR1}). This also happens for the hottest stars in IC2714 (lower right panel of Fig.\,\ref{HR1}), NGC2539, NGC6633, and NGC5822 (see Fig.\,\ref{HR2} for the HR diagrams of these
clusters). On the other hand, other stars arrive at the base of the RGB with undetectable Li abundances, probably because they have completed the FDU dilution (see the triangle symbols in NGC4349, NGC2360, or NGC5822). Lithium depleted stars can be found in other evolutionary stages as well. For example, the three stars in NGC2682 seem to be in the so-called red clump where they have just started the core He-burning. This might be the case for the coolest star in NGC2539 and several objects in IC4651.

\subsection{Li-enhanced stars}

We now focus on the coolest objects that show significantly strong Li lines. These objects are IC4651No9791, NGC2423No3, and NGC4349No127 (with Li EWs higher than 100\,m$\AA{}$). For a comparison of the spectra of these stars see Figs.\,\ref{IC4651}, \ref{NGC2423}, and \ref{NGC4349}, respectively, in  Appendix \ref{app:plots}. The stars NGC3680No13 and NGC3680No26 also show relatively strong EWs, close to 80\,m$\AA{}$ (Fig.\,\ref{NGC3680}).
\vspace*{0.3cm}
\\* \textit{IC4651No9791}\linebreak
In this cluster most of the stars have completely destroyed their Li (only showing upper limits) except IC4651No9791 (\teff\,=\,4537K, \logg\,=\,2.44\,dex), which shows a higher Li abundance (A(Li)\,=\,1.14\,dex, EW\,=\,106.7m\,$\AA{}$, A(Li)$_{NLTE}$\,=\,1.41\,dex) than other cluster members. The isochrone that most closely matches our stars is quite close to the age estimate given by the WEBDA database. The masses of these stars are $\sim$\,2.1\,M$_\odot$, thus they may experience the luminosity bump. Indeed, the most Li-rich star in this cluster seems to be close to this evolutionary phase where the erasing of the chemical discontinuity left behind by the FDU might allow some extra-mixing to trigger the Cool Bottom Process \citep{sackmann99}. We note here that the maximum allowed mass to experience the luminosity bump depends on the metallicity and the isochrone chosen here does not show this stage.
Many of the discovered Li-rich stars are found to be close to the luminosity bump, as shown by CB00 and \citet{kumar11}. However, being in the luminosity bump does not guarantee the appearance of Li-rich stars. As explained in CB00, the efficiency of mixing will determine whether Li reaches the surface or not. Once it appears, it might not survive a very long time: when the mixing extends deep enough to convert additional $^{12}$C into $^{13}$C (as is observed to occur in low-mass giants) the temperatures get very hot and Li is burned again. Therefore, this star might be in a Li enrichment process (its abundance is not as high as observed in other Li-rich stars) or might already be starting to destroy the Li created during the luminosity bump. 
\vspace*{0.3cm}
\\* \textit{NGC4349No127}\linebreak
This star is the coolest (\teff\,=\,4503K) and most evolved (\logg\,=\,1.86\,dex) in the cluster and shows a strong Li absorption (118\,m$\AA{}$) line with A(Li)\,=\,1.1\,dex (A(Li)$_{NLTE}$\,=\,1.42\,dex). Its mass is certainly higher than 2-2.3\,M$_\odot$, so this star cannot experience the luminosity bump because the RGB finishes before the hydrogen shell reaches the chemical discontinuity and thus no Cool Bottom Process is expected \citep{sackmann99}. It cannot have experienced the He-flash either. Alternatively, the Hot Bottom Burning mechanism can be activated during the He-shell flashes on the AGB for stars more massive than 4\,M$_\odot$ \citep{sackmann92}. The derived mass for this star is 3.8M$_\odot$ and it seems to be either on the first or second ascent of the RGB, hence neither of the previous mechanisms is able to explain its enhanced Li.
This star has a logarithmic luminosity of 2.81\,L$_\odot$, close to the values of the Li-rich stars on the early AGB compiled by CB00. In that work the authors suggest that some mechanism must act at this evolutionary stage to trigger the CFM. From its position in the HR diagram it is not clear whether this star has already reached the early AGB.

Moreover, a similar cluster in age and metallicity, NGC3114No181 (\teff\,=\,4543K, \logg\,=\,1.86\,dex), has very similar parameters to NGC4349No127 and a much lower Li abundance (A(Li)\,$<$\,0.15\,dex). Interestingly, NGC4349No127 hosts a brown dwarf of 19.8\,M$_J$ at a distance of 2.4\,AU with a period of 678 days (LM07). One can then speculate that a possible explanation for the enhanced Li observed in this star is the engulfment of a putative short-period planet that previously existed in the system as might be the case of other Li-rich planet hosts presented in \citet{adamow12,adamow14}. 
\vspace*{0.3cm}
\\* \textit{NGC2423No3}\linebreak
This star belongs to a cluster for which we only have the spectra of three stars, making the selection of the more suitable isochrones more difficult. The ages that most closely match our stars are older than the literature values and predict a mass for our stars close to $\sim$\,2\,M$_\odot$, as in IC4651. In this cluster, the most Li-rich star is the coolest and most evolved (\teff\,=\,4592K, \logg\,=\,2.31\,dex); it also has a strong Li line (107\,m$\AA{}$), A(Li)\,=1.25\,dex (A(Li)$_{NLTE}$\,=\,1.52\,dex). This star has very similar parameters to IC4651No9791 and also seems to be close to the luminosity bump, which would explain its higher Li content. However, if the age of this cluster is really younger, then the stars at the RGB would be more massive (2.4\,M$_\odot$) and in principle would not experience the luminosity bump. Therefore, another mixing mechanism might be necessary to explain its higher abundance. In Fig.\,\ref{HR1} we also depict a younger isochrone (dashed line, 0.9Gyr) for this cluster that more closely matches the hotter stars in the sample.
We note that NGC2423No3 hosts a giant planet (10.6\,M$_J$) at a distance of 2.1\,AU and a period of 714 days (LM07). Then, as for NGC4349No127, there might be a connection between the Li enhancement and the accretion of planetary material. Certainly, the determination of the $^{12}$C/$^{13}$C would help to constrain the evolutionary state of our stars (pre- or post-luminosity bump) and to distinguish between an internal production of Li (where extra-mixing is needed and hence $^{12}$C/$^{13}$C decreases) or an external contamination due to planet engulfment where an increase of $^{12}$C/$^{13}$C is expected \citep{carlberg12}. Unfortunately, the CN lines suitable for the determination of this ratio, located around 8000\,$\AA{}$, are not covered by our spectra.
\vspace*{0.3cm}
\\* \textit{NGC3680No13/No26}\linebreak
This cluster seems to be the best example of stars exactly positioned at the luminosity bump (see the two stars with almost identical parameters in lower left panel of Fig.\,\ref{HR1}). Despite the lack of warmer stars, we can match our stars with an isochrone that agrees very well with the age given by WEBDA. The three objects in the cluster show detectable Li lines and the two coolest stars present higher abundances (see Fig.\,\ref{NGC3680}, NLTE Li abundances of 1.44 and 1.34\,dex), which might be a sign of the beginning of Li enrichment during the luminosity bump.

\begin{figure}
\centering
\includegraphics[width=1.0\linewidth]{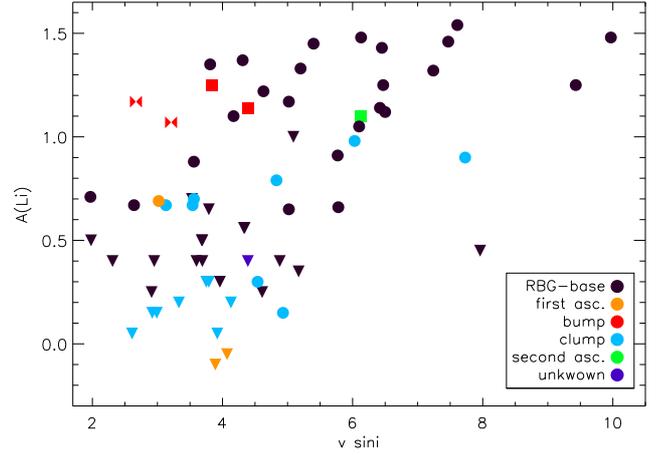}
\caption{Li vs. projected rotational velocities. The symbols are the same as in Fig.\,\ref{HR1} and the colours denote the estimated evolutionary stage.} 
\label{Livsini}
\end{figure}

\subsection{Effect of planet engulfment?}
If the cause of the Li enhancement in these stars were the engulfment of a planet we could expect a higher rotation rate due to the transfer of angular momentum and an increase in elements that are important for the formation of planets \citep{denissenkov04,carlberg10,carlberg12}.
In order to check whether the stars with higher Li show a different abundance pattern from their counterparts in the same cluster, we have derived abundances for 12 refractory elements ($\alpha$ and Fe-peak), in the same way as \citet{adibekyan15}, which will be presented in a separate work (Zaworska et al. in prep.). All the stars within the same cluster show similar abundances within the errors, and no clear enhancement in a given element is observed for any of the Li-enhanced stars.\\

We also derived the rotational projected velocities (\textit{v} sin \textit{i}) by fitting several atomic lines with spectral synthesis using the software Spectroscopy Made Easy \citep{valenti96}. For further details about this determination we refer the reader to \citet{tsantaki14}.
In Fig.\,\ref{Livsini} we show the \textit{v} sin \textit{i} values for our stars divided into several evolutionary stages as shown in the legend (this division has been made by eye using Figs. \ref{HR1} and \ref{HR2}; see the assigned stages in Table \ref{parameters}). 
We note that this selection is subjective and for some clusters it is not possible to distinguish among different stages, for example between the first ascent or the red clump. A further explanation of the evolutionary stages for our stars can be found in Sect.\,\ref{app:iso}. We see a clear increasing trend of A(Li) with \textit{v} sin \textit{i}. A similar trend has been observed by \cite{carlberg12}; the authors find that rapid rotators (\textit{v} sin \textit{i}\,$>$\,8 \kms) are on average enriched in lithium when compared to the slow rotators, regardless of the evolutionary stage. In our sample, for example, among the RGB base, the stars with higher Li abundances (black circles) are shifted towards higher rotation velocities than Li-depleted stars (black triangles). Nevertheless, our figure might simply be a reflection of the expected higher rotation rates for hotter stars (most of the stars with high Li are also the hottest). In our sample we find that the Li-enhanced stars belonging to the bump show similar \textit{v} sin \textit{i} values (2.5\,-\,4.5\,\kms) to the stars with no detected Li. Only NGC4349No127 (green square in Fig.\,\ref{Livsini}) shows a higher \textit{v} sin \textit{i}. This is the coolest star (more than 250K difference) among the stars with \textit{v} sin \textit{i}\,$>$\,5 \kms\ and it also rotates faster than the most evolved stars within its cluster. A possible explanation for its enhanced Li and rotation might be the engulfment of a planetary body. If this were true, the engulfment should have happened some time ago since the rotation rate is not very fast and thus the star has had time to slow down. Moreover, a putative engulfed planet should have been at a very short orbit, since for stars with M\,$\sim$\,3\,M$_\odot$ the critical initial semi-major axis for engulfment is $\sim$\,0.2-0.3\,AU at the RGB phase \citep{villaver09,kunitomo11}.\\

\begin{figure}
  \centering
  \includegraphics[width=1.0\linewidth]{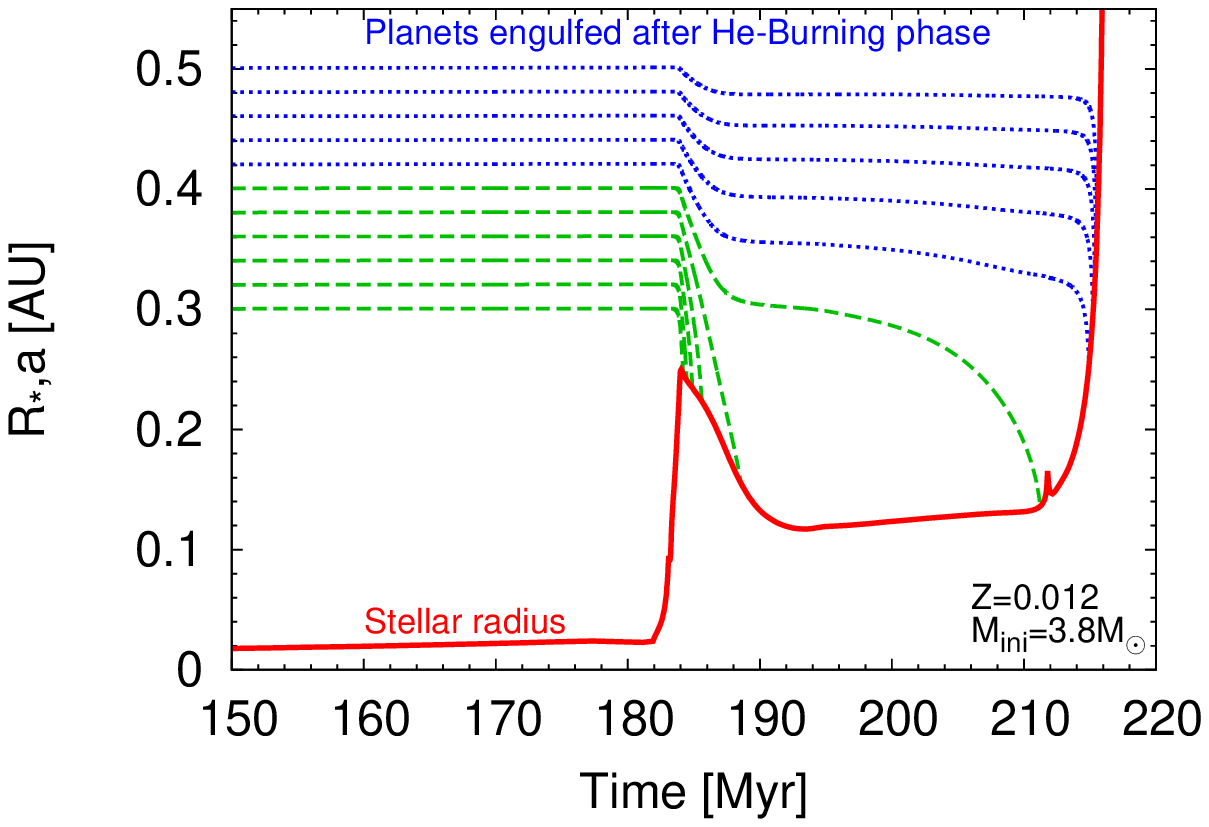}
  \includegraphics[width=1.0\linewidth]{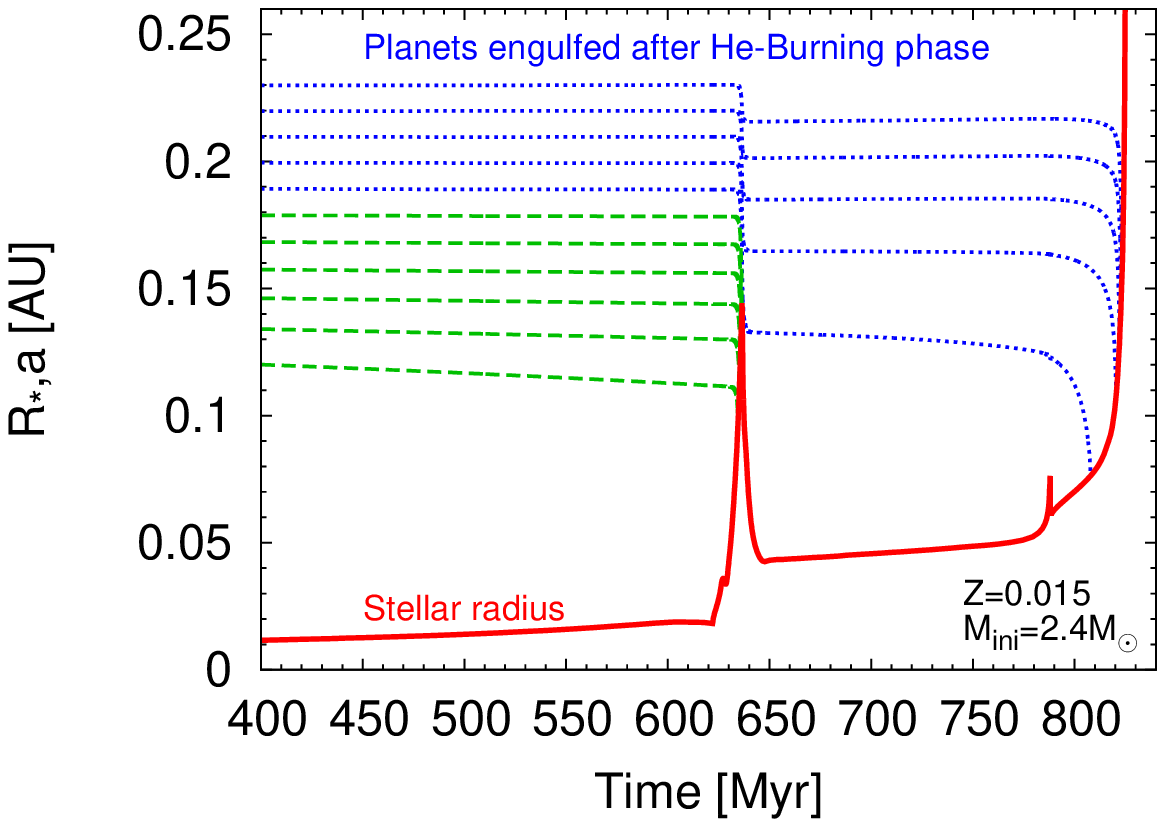}
  \caption{Orbital evolution of 1\,M$_J$ planets with different initial semi-major axes around NGC4349No127 (upper panel) and NGC2423No3 (lower panel). The green dashed lines show planets that will be engulfed during RGB, while blue pointed lines show the tracks of planets engulfed after the He-burning phase. The evolution of stellar radius is shown with a red line.}
  \label{evolution}
  \end{figure}

To further investigate this possibility we calculate the orbital evolution of hypothetical planets around NGC4349No127 and NGC2423No3 using the stellar evolution code MESA \citep{paxton11} as was done in \citet{kunitomo11}. The evolution starts at zero-age main sequence (ZAMS) and ends before the star experiences the thermal pulses on the AGB (see Fig.\,\ref{evolution}). For NGC4349No127 we have considered our derived metallicity and the mass we obtain from PARSEC isochrones. The isochrones give an age of 203\,$\pm$31\,Myr for this star, close to the literature value. Therefore, this star would be in the He-burning phase and it could not have engulfed a planet during the last $\sim$\,20\,Myr, once the stellar radius was reduced after the high increase taking place at the RGB tip (at $\sim$\,185\,Myr in the top panel of Fig.\,\ref{evolution}). There is only a small probability of engulfment if a planet with an initial semi-major axis of about 0.4\,AU existed in the system (see the green track reaching the stellar radius at $\sim$\,212\,Myr). Nevertheless, considering the error on the age, this star might be either on the first or the second ascent in the RGB. The evolutionary status of this star is also not clear from what we observe in the HR diagram of Fig.\,\ref{HR1}. If this star is still ascending the RGB for the first time, then any planet with semi-major axis lower than 0.38\,AU would be engulfed close to the RGB tip. The question remains whether a possible enhancement of Li caused by this process can survive during the He-burning phase. Since the Li abundance observed in this star is not very high, it is possible to speculate that this star was a Li-rich giant in the past and now we are observing the remaining Li after the RGB tip. On the other hand, if this star is ending the core He-burning phase (i.e. second ascent) and Li cannot survive after the RGB tip, its enhanced Li should be caused by a mechanism other than planet engulfment. The Cameron-Fowler mechanism would also not work because the convective envelope is very shallow during the He-burning phase and thus any mixing process would not be able to connect the cool convective envelope to the hotter regions where $^7$Be can be synthesized. Finally, if this star is more evolved than estimated (R\,$\gtrsim$\,0.3\,AU), it will be able to engulf planets at a larger semi-major axis during the early AGB when the star expands significantly and its convective envelope deepens (at ages $>$\,215\,Myr). We note that it is at this phase, before the second dredge-up, that CB00 also find several stars with mild Li abundances. They suggest that an unknown mixing episode happening at this phase may connect the deeper convective envelope to the $^3$He rich shell to trigger the CFM. Therefore, the probable engulfment of close-in planets during the early AGB (as shown in Fig.\,\ref{evolution}) might cause the needed extra-mixing owing to the shear instabilities triggered by the dissipation processes after planet engulfment \citep{siess99}.\\

For NGC2423No3 we have considered the possibility that our derived mass and age are not correct since we obtain older ages than in the literature. Our derived mass for this star (2.2\,M$_\odot$) falls close to the boundary where the luminosity bump is experienced (at Z\,=\,0.018 the luminosity bump occurs for M\,$<$\,2\,M$_\odot$ in PARSEC isochrones, while for Z\,=\,0.015 it happens for M\,$<$\,1.9\,M$_\odot$), where -- as previously discussed -- the Cool Bottom Process can explain the enhanced Li. However, for the literature age of 740\,Myr we would obtain a mass of 2.4\,M$_\odot$, which implies a low probability of encountering the bump. In that case, this star would be in a similar situation to NGC4349No127, which is undergoing the He-burning phase. The probability of engulfing a planet at this phase is negligible; thus, its enhanced Li must be a residue of an earlier enrichment event, considering that Li cannot be produced at this evolutionary stage. Otherwise, this star might be at a more evolved phase (R\,$\gtrsim$\,0.3\,AU), in the early AGB where planets at short orbits will barely survive.

\section{Conclusions} \label{sec:conclusions}
We present Li abundances for a sample of 67 red giants in 12 different open clusters. We find that five objects show enhanced Li abundances and interestingly, the only two stars with detected substellar companions in the sample are among them. Four out of these five Li-enhanced stars seem to be very close to the luminosity bump experienced by low-mass red giants. This phase has already been proposed as a probable scenario for the appearance of Li-rich stars \citep[CB00,][]{kumar11,sackmann99}. However, NGC4349No127, a star hosting a brown dwarf, seems to be in a more evolved phase where it is not clear how an extra mixing process might complement the CFM. We note that this star also seems to be far from the early AGB phase, which was pointed out as another phase for Li enrichment by CB00. The connection between the Li enhancement for this star and a planetary engulfment scenario sounds appealing, but it is difficult to probe. The orbital evolution of hypothetical Jupiter-type planets around this star only predicts an engulfment episode for planets at 0.4\,AU during the He-burning phase. However, if we consider the errors in age and mass, this star might be at another evolutionary stage where the engulfment of planets is more probable. We suggest that the plausible accretion of planets (and the triggered instabilities) on the early AGB (see Fig.\,\ref{evolution}) might be one of the causes to trigger the needed extra-mixing in order to produce Li enhancement due to the CFM, as observed in other early AGB stars. We find that this might also be the case for NGC2423No3, the other star hosting a substellar companion in our sample.
A closer inspection of these stars might help to discard or support this scenario. We plan to obtain better spectra for some of our stars in order to measure the $^{6}$Li/$^{7}$Li ratio since the presence of $^{6}$Li in the photosphere of a giant star would support the engulfment scenario \citep[e.g.][]{israelian03}. The determination of the $^{12}$C/$^{13}$C ratio and the Be abundance would certainly be very useful. Several works have failed to find Be in Li-rich giants; this element should be detected if the Li enrichment originates from a planet engulfment \citep[e.g.][]{melo05}. Furthermore, the tendency of Li-rich stars to accumulate in specific evolutionary phases does not support this scenario since Li-rich giants produced by planetary accretion should appear across the HR diagram. Nevertheless, a few cases of Li-rich giants have been found in evolutionary phases where Li internal production is not expected \citep[e.g.][]{alcala11,carlberg10,carlberg15}. Interestingly, the probability of engulfing close-in planets is high on the RGB tip and on the early AGB, just two of the evolutionary stages where many Li-rich stars are found, but where the origin of an extra-mixing episode is not clear (unlike the well-known mixing that takes place during the luminosity bump). Thus, we can speculate that the engulfment of planets might be the triggering episode for extra-mixing for some of these stars, which would in turn produce Li through the CFM rather than directly accreting it from a hosting planet. A similar mechanism was proposed by \citet{denissenkov04}, though the transfer of angular momentum from a planet should be noticeable on the rotational velocity of the star. These authors report that a factor of $\sim$\,10\, increase in the giant's spin angular velocity is needed to account for the needed mixing. The discovery of new Li-rich stars and the current surveys that are searching for planets in open clusters might help to shed some light on this issue. 

\begin{acknowledgements}
E.D.M and V.Zh.A. acknowledge the support from the Funda\c{c}\~ao para a Ci\^encia e Tecnologia, FCT (Portugal) in the form of the grants SFRH/BPD/76606/2011 and SFRH/BPD/70574/2010, respectively. N.C.S. and S.G.S. also acknowledge the support from FCT through Investigador FCT contracts of reference IF/00169/2012 and IF/00028/2014, respectively, and POPH/FSE (EC) by FEDER funding through the program ``Programa Operacional de Factores de Competitividade - COMPETE''. M.K. is supported by MEXT of Japan (Grant: 23244027). This work results within the collaboration of the COST Action TD 1308. 
This research has made use of the SIMBAD and WEBDA databases. This work has also made use of the IRAF facility.

\end{acknowledgements}
\vspace{-0.4cm}

\bibliography{edm_bibliography}
\clearpage

\appendix

\section{Detailed HR diagram}  \label{app:iso}

\begin{center}
\begin{table}[h]
\caption{Minimum and maximum radius reached at each evolutionary phase depicted in Fig.\,\ref{example} for the older isochrone (upper table) and younger isochrone (lower part of the table).}
\centering
\begin{tabular}{lcccc}
\hline
\noalign{\smallskip} 
Red clump mass      & 1.8\,M$_\odot$ &    &   3.7\,M$_\odot$ &   \\ 
\noalign{\smallskip} 
\hline
\noalign{\smallskip} 
Phase & R$_{min}$ & R$_{max}$ & R$_{min}$ & R$_{max}$  \\  
     & (R$_\odot$)  &  (R$_\odot$) & (R$_\odot$)  &  (R$_\odot$) \\
\noalign{\smallskip} 
\hline
\noalign{\smallskip} 
RGB base                   &  4.2  &    4.9    &  19.7   &  24.8  \\
first ascent pre-bump      &  4.9  &   10.4    &  24.8   &  51.1  \\
luminosity bump            &  10.4  &  22.0    &  ---    &  ---  \\
first ascent-post bump     &  22.0  &  62.5    &  ---    &  ---  \\
RGB tip                    &  62.5  &  95.6    &  51.1   &  63.1  \\
red clump                  &  9.8  &   17.2    &  27.0   &  42.9   \\
second ascent              &  17.2  &  113.0   &  42.9   &  346.6  \\
\hline
\noalign{\medskip} %
\end{tabular}
\end{table}
\label{radius}
\end{center}

\vspace{-0.2cm}

\begin{figure}
\centering
\includegraphics[width=9.0cm]{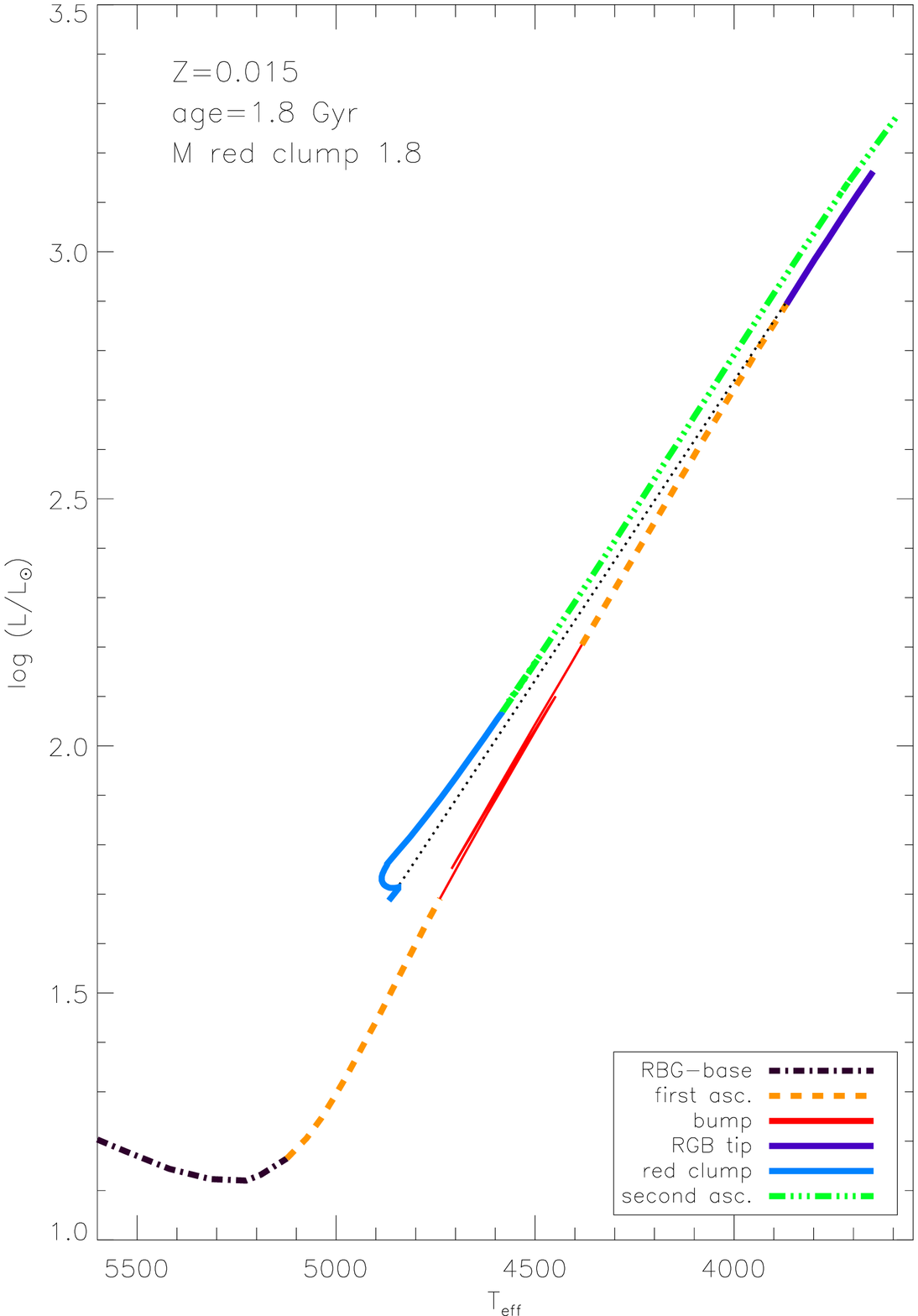}
\includegraphics[width=9.0cm]{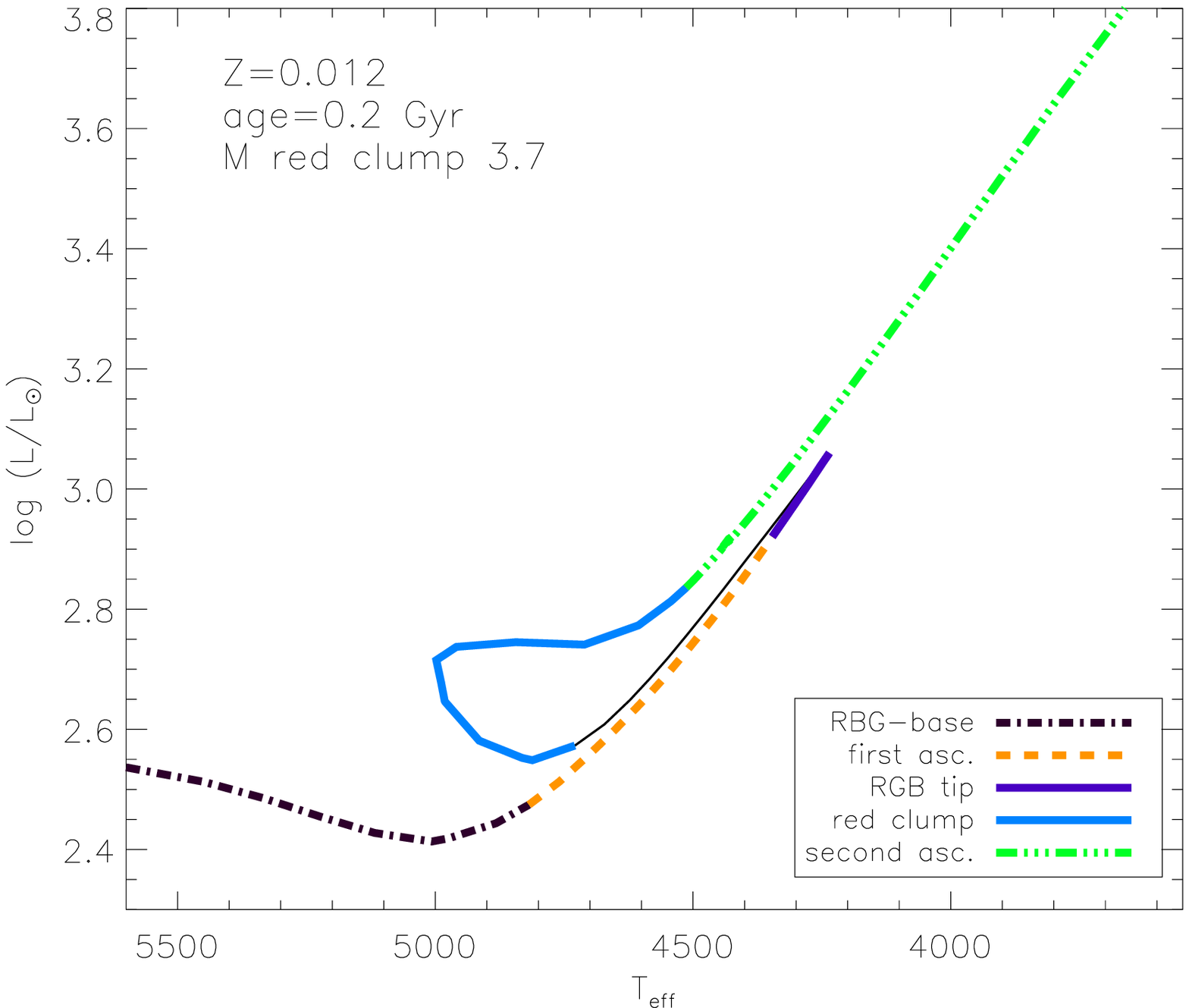}
\caption{Schematic HR diagram for stars at two different ages and metallicities.} 
\label{example}
\end{figure}

This section gives a schematic explanation of the evolution on a HR diagram to serve as a guide for the evolutionary stages assigned to our stars in Table \ref{parameters}. We note that the determination of these stages has been made in a qualitative way and the boundaries between them are therefore approximate. In Fig.\,\ref{example} we show an isochrone of $\sim$\,1.8\,Gyr for solar metallicity \citep{bressan12} where the different evolutionary phases are indicated with different colours. For reference, a star with that age in the red clump will have a mass of 1.77\,M$_\odot$. Many of our stars lie on the base of the RGB where the star has already started its expansion (R\,$\sim$\,5\,R$_\odot$, see Table \ref{radius}), it is quietly burning H in a shell around the core, and Li has already experienced an important dilution due to the FDU. Once the star begins its first ascent in the giant branch, the FDU is close to finishing (see also Fig.\,1 of CB00). At this phase the radius can reach values of R\,$\sim$\,10\,R$_\odot$ prior to the luminosity bump. During the bump, the chemical discontinuity is erased by the outward H-burning shell allowing for any extra-mixing to connect this shell to the convective envelope. In the HR diagram this is shown as a brief phase of decreasing luminosity. The maximum radius at this phase is R\,$\sim$\,22\,R$_\odot$. After the bump, the star continues its ascent until the RGB tip, the point of maximum luminosity where the radius reaches a local maximum (R\,$\sim$\,95\,R$_\odot$). It is in this phase that close-in planets can be engulfed (see the first sudden increase of radius in Fig.\,\ref{evolution}). For stars less massive than $\sim$\,2\,M$_\odot$ the He core becomes degenerate and its ignition is called the He-flash. Here there is a discontinuity in the isochrone where the star gets hotter and significantly reduces its luminosity and radius (from $\sim$\,95\,R$_\odot$ to $\sim$\,10\,R$_\odot$) until it starts to burn He in the core. This is the red clump phase, considered the metal-rich counterpart to the horizontal branch. The timescale for He-core burning is longer than previous phases, thus it is easier to observe stars at this evolutionary stage, producing a clump of stars on the HR diagram. The core He exhaustion marks the end of the red clump phase and the star begins its second ascent in the giant branch towards the early AGB, where it will expand again to reach R$>$\,100\,R$_\odot$. Close-in planets will barely survive this phase of fast expansion by their hosting stars (see Fig.\,\ref{evolution}). In Fig.\,\ref{example} we also show the different phases for an isochrone of 200\,Myr when the stars at the RGB will have masses close to 3.7\,M$_\odot$. The global picture for the intermediate-mass stars of our sample (2\,$<$R/R$_\odot$\,$<$\,4) is similar to the previous case, but the evolution will be faster, thus those stars will not experience the luminosity bump. They will reach hotter temperatures in the core so the He will not become degenerate and they will not experience the He-flash. As a consequence, the maximum radius attained at the RGB tip is smaller than for less massive stars and the probability of engulfing a planet is lower.

\section{Stellar parameters and abundances}\label{app:tables}

\begin{center}
\begin{table*}[h]
\caption{Stellar spectroscopic parameters and abundances for our sample}
\centering
\begin{tabular}{lcccrccccccc}
\hline
\noalign{\medskip} 
Star & $T_\mathrm{eff}$ & $\log{g}$ & $\xi_t$ & [Fe/H] & ${\rm A(Li)}$ & NLTE A(Li) & \textit{v} sin \textit{i} & Evol. stage & Instrument\\  
     & (K)  &  (cm\,s$^{-2}$) & (km\,s$^{-1}$) &  &  & &  (\kms)&  &  \\
\noalign{\medskip} 
\hline
\hline
    NGC3114No262  &   4906 $\pm$  93  &  2.66 $\pm$ 0.18  &  2.09 $\pm$ 0.11  & -0.12 $\pm$ 0.07  &     1.25 $\pm$  0.18  &  1.44    &  9.43  &   rgb-base  &   HARPS  \\
      NGC3114No6  &   4751 $\pm$  97  &  1.93 $\pm$ 0.19  &  2.47 $\pm$ 0.14  & -0.15 $\pm$ 0.08  &     0.90 $\pm$  0.14  &  1.16    &  7.73  &   clump     &   HARPS  \\
     NGC3114No13  &   4912 $\pm$  39  &  2.62 $\pm$ 0.10  &  1.54 $\pm$ 0.04  & -0.05 $\pm$ 0.03  &  $<$0.25              &  0.45    &  2.91  &   rgb-base  &   UVES  \\
    NGC3114No181  &   4543 $\pm$  96  &  1.90 $\pm$ 0.22  &  1.94 $\pm$ 0.10  & -0.12 $\pm$ 0.06  &     0.15 $\pm$  0.19  &  0.47    &  4.93  &   clump     &   UVES  \\
    NGC3114No273  &   5278 $\pm$  28  &  2.95 $\pm$ 0.07  &  1.57 $\pm$ 0.03  &  0.05 $\pm$ 0.03  &  $<$0.40              &  0.56    &  2.31  &   rgb-base  &   UVES  \\
\hline
    NGC4349No127  &   4503 $\pm$  70  &  1.99 $\pm$ 0.19  &  1.77 $\pm$ 0.07  & -0.13 $\pm$ 0.04  &     1.10 $\pm$  0.14  &  1.42    &  6.13  &     second  &   HARPS  \\
    NGC4349No168  &   5186 $\pm$  59  &  2.93 $\pm$ 0.11  &  1.95 $\pm$ 0.07  & -0.01 $\pm$ 0.05  &     1.12 $\pm$  0.22  &  1.29    &  6.50  &   rgb-base  &   HARPS  \\
    NGC4349No174  &   4676 $\pm$  87  &  2.25 $\pm$ 0.19  &  1.85 $\pm$ 0.09  & -0.15 $\pm$ 0.06  &     0.79 $\pm$  0.12  &  1.11    &  4.83  &     clump   &   HARPS  \\
    NGC4349No203  &   5271 $\pm$  71  &  3.35 $\pm$ 0.13  &  2.01 $\pm$ 0.09  &  0.04 $\pm$ 0.06  &     1.25 $\pm$  0.13  &  1.38    &  6.47  &   rgb-base  &   HARPS  \\
      NGC4349No5  &   5166 $\pm$  87  &  3.16 $\pm$ 0.18  &  2.24 $\pm$ 0.13  & -0.08 $\pm$ 0.07  &  $<$0.45              &  0.63    &  7.96  &   rgb-base  &   HARPS  \\
     NGC4349No53  &   4880 $\pm$  40  &  2.66 $\pm$ 0.10  &  1.49 $\pm$ 0.04  & -0.11 $\pm$ 0.03  &  $<$0.30              &  0.50    &  3.79  &      clump  &   HARPS  \\
      NGC4349No9  &   5382 $\pm$ 136  &  3.95 $\pm$ 0.22  &  2.52 $\pm$ 0.28  & -0.08 $\pm$ 0.10  &     1.48 $\pm$  0.18  &  1.58    &  9.97  &   rgb-base  &   HARPS  \\
\hline
     IC2714No110  &   5145 $\pm$  59  &  3.03 $\pm$ 0.10  &  1.54 $\pm$ 0.06  &  0.09 $\pm$ 0.05  &     0.66 $\pm$  0.21  &  0.83    &  5.78  &   rgb-base  &   HARPS  \\
     IC2714No121  &   4768 $\pm$  49  &  2.34 $\pm$ 0.12  &  1.77 $\pm$ 0.05  & -0.04 $\pm$ 0.04  &     0.30 $\pm$  0.22  &  0.53    &  4.54  &      clump  &   HARPS  \\
     IC2714No126  &   5211 $\pm$  77  &  3.21 $\pm$ 0.22  &  2.12 $\pm$ 0.10  &  0.07 $\pm$ 0.06  &     1.46 $\pm$  0.12  &  1.60    &  7.47  &   rgb-base  &   HARPS  \\
     IC2714No190  &   5077 $\pm$  86  &  2.82 $\pm$ 0.18  &  1.91 $\pm$ 0.10  &  0.00 $\pm$ 0.07  &     1.05 $\pm$  0.18  &  1.23    &  6.10  &   rgb-base  &   HARPS  \\
     IC2714No220  &   4980 $\pm$  64  &  2.57 $\pm$ 0.14  &  1.75 $\pm$ 0.07  & -0.01 $\pm$ 0.05  &     0.98 $\pm$  0.17  &  1.16    &  6.03  &      clump  &   HARPS  \\
      IC2714No53  &   5261 $\pm$  64  &  3.01 $\pm$ 0.17  &  1.68 $\pm$ 0.07  &  0.08 $\pm$ 0.06  &     1.43 $\pm$  0.13  &  1.56    &  6.45  &   rgb-base  &   HARPS  \\
       IC2714No5  &   5224 $\pm$  53  &  2.85 $\pm$ 0.17  &  1.91 $\pm$ 0.07  &  0.04 $\pm$ 0.05  &     1.32 $\pm$  0.13  &  1.46    &  7.24  &   rgb-base  &   HARPS  \\
      IC2714No87  &   5377 $\pm$  70  &  3.45 $\pm$ 0.17  &  1.69 $\pm$ 0.08  &  0.11 $\pm$ 0.06  &     1.54 $\pm$  0.18  &  1.63    &  7.61  &   rgb-base  &   HARPS  \\
\hline
    NGC2539No229  &   5157 $\pm$  38  &  3.06 $\pm$ 0.09  &  1.44 $\pm$ 0.04  &  0.12 $\pm$ 0.03  &     1.35 $\pm$  0.12  &  1.50    &  3.81  &   rgb-base  &   HARPS  \\
    NGC2539No246  &   5176 $\pm$  43  &  3.00 $\pm$ 0.10  &  1.50 $\pm$ 0.04  &  0.08 $\pm$ 0.04  &     1.14 $\pm$  0.17  &  1.29    &  6.42  &   rgb-base  &   HARPS  \\
    NGC2539No251  &   5168 $\pm$  36  &  3.02 $\pm$ 0.09  &  1.39 $\pm$ 0.04  &  0.07 $\pm$ 0.03  &     0.91 $\pm$  0.11  &  1.07    &  5.77  &   rgb-base  &   HARPS  \\
    NGC2539No317  &   5149 $\pm$  42  &  2.86 $\pm$ 0.08  &  1.56 $\pm$ 0.05  &  0.08 $\pm$ 0.04  &     1.17 $\pm$  0.12  &  1.34    &  5.02  &   rgb-base  &   HARPS  \\
    NGC2539No346  &   5104 $\pm$  41  &  2.78 $\pm$ 0.08  &  1.52 $\pm$ 0.04  &  0.03 $\pm$ 0.04  &     1.37 $\pm$  0.11  &  1.54    &  4.31  &   rgb-base  &   HARPS  \\
    NGC2539No447  &   5030 $\pm$  41  &  2.66 $\pm$ 0.10  &  1.60 $\pm$ 0.04  &  0.08 $\pm$ 0.04  &  $<$0.25              &  0.42    &  4.61  &   rgb-base  &   HARPS  \\
    NGC2539No463  &   5046 $\pm$  50  &  2.79 $\pm$ 0.10  &  1.62 $\pm$ 0.05  &  0.06 $\pm$ 0.04  &     0.65 $\pm$  0.16  &  0.82    &  5.02  &   rgb-base  &   HARPS  \\
    NGC2539No502  &   5166 $\pm$  36  &  2.91 $\pm$ 0.06  &  1.52 $\pm$ 0.04  &  0.08 $\pm$ 0.03  &  $<$0.35              &  0.52    &  5.17  &   rgb-base  &   HARPS  \\
    NGC2539No652  &   4813 $\pm$  47  &  2.60 $\pm$ 0.12  &  1.52 $\pm$ 0.04  &  0.02 $\pm$ 0.03  &  $<$0.05              &  0.22    &  3.92  &      clump  &   HARPS  \\
\hline
    NGC6633No106  &   5199 $\pm$  42  &  2.97 $\pm$ 0.09  &  1.50 $\pm$ 0.05  &  0.03 $\pm$ 0.04  &     1.48 $\pm$  0.13  &  1.63    &  6.13  &   rgb-base  &   HARPS  \\
    NGC6633No100  &   5071 $\pm$  34  &  2.70 $\pm$ 0.09  &  1.60 $\pm$ 0.04  &  0.00 $\pm$ 0.03  &     0.88 $\pm$  0.16  &  1.03    &  3.56  &   rgb-base  &   UVES  \\
    NGC6633No119  &   5246 $\pm$  31  &  3.04 $\pm$ 0.18  &  1.41 $\pm$ 0.04  & -0.01 $\pm$ 0.03  &     1.33 $\pm$  0.14  &  1.45    &  5.20  &   rgb-base  &   UVES  \\
    NGC6633No126  &   5217 $\pm$  36  &  3.09 $\pm$ 0.15  &  1.47 $\pm$ 0.04  &  0.02 $\pm$ 0.03  &     1.22 $\pm$  0.18  &  1.34    &  4.63  &   rgb-base  &   UVES  \\
\hline
      IC4756No44  &   5266 $\pm$  62  &  3.31 $\pm$ 0.16  &  1.41 $\pm$ 0.07  &  0.18 $\pm$ 0.05  &  $<$1.00              &  1.12    &  5.09  &   rgb-base  &   HARPS  \\
     IC4756No125  &   5172 $\pm$  30  &  3.00 $\pm$ 0.08  &  1.44 $\pm$ 0.03  & -0.01 $\pm$ 0.03  &  $<$0.50              &  0.65    &  1.98  &   rgb-base  &   UVES  \\
      IC4756No38  &   5170 $\pm$  25  &  3.01 $\pm$ 0.06  &  1.36 $\pm$ 0.03  &  0.00 $\pm$ 0.02  &     0.71 $\pm$  0.16  &  0.87    &  1.97  &   rgb-base  &   UVES  \\
      IC4756No42  &   5249 $\pm$  31  &  3.11 $\pm$ 0.07  &  1.38 $\pm$ 0.04  &  0.02 $\pm$ 0.03  &     0.67 $\pm$  0.22  &  0.79    &  2.64  &   rgb-base  &   UVES  \\
\hline
    NGC2360No119  &   5160 $\pm$  36  &  3.03 $\pm$ 0.06  &  1.50 $\pm$ 0.04  &  0.01 $\pm$ 0.03  &  $<$0.50              &  0.67    &  3.69  &   rgb-base  &   HARPS  \\
     NGC2360No50  &   5162 $\pm$  40  &  3.05 $\pm$ 0.06  &  1.47 $\pm$ 0.04  &  0.01 $\pm$ 0.04  &  $<$0.56              &  0.72    &  4.34  &   rgb-base  &   HARPS  \\
     NGC2360No66  &   5309 $\pm$  33  &  2.97 $\pm$ 0.28  &  1.37 $\pm$ 0.04  &  0.02 $\pm$ 0.03  &  $<$0.40              &  0.48    &  4.88  &   rgb-base  &   HARPS  \\
     NGC2360No79  &   5167 $\pm$  31  &  3.17 $\pm$ 0.07  &  1.37 $\pm$ 0.03  &  0.03 $\pm$ 0.03  &  $<$0.40              &  0.56    &  3.60  &   rgb-base  &   HARPS  \\
      NGC2360No7  &   5189 $\pm$  31  &  3.12 $\pm$ 0.06  &  1.47 $\pm$ 0.04  &  0.01 $\pm$ 0.03  &  $<$0.50              &  0.65    &  3.68  &   rgb-base  &   HARPS  \\
     NGC2360No85  &   5193 $\pm$  32  &  3.12 $\pm$ 0.09  &  1.39 $\pm$ 0.03  &  0.01 $\pm$ 0.03  &  $<$0.70              &  0.86    &  3.53  &   rgb-base  &   HARPS  \\
     NGC2360No86  &   5052 $\pm$  32  &  2.90 $\pm$ 0.10  &  1.36 $\pm$ 0.03  & -0.03 $\pm$ 0.03  &  $<$0.30              &  0.47    &  3.96  &   rgb-base  &   HARPS  \\
     NGC2360No89  &   5151 $\pm$  34  &  2.97 $\pm$ 0.11  &  1.46 $\pm$ 0.04  & -0.02 $\pm$ 0.03  &  $<$0.56              &  0.73    &  4.33  &   rgb-base  &   HARPS  \\
\hline
    NGC5822No201  &   5347 $\pm$  58  &  3.07 $\pm$ 0.09  &  1.54 $\pm$ 0.06  &  0.06 $\pm$ 0.05  &     1.10 $\pm$  0.22  &  1.18    &  4.17  &   rgb-base  &   HARPS  \\
      NGC5822No8  &   5124 $\pm$  34  &  3.01 $\pm$ 0.06  &  1.50 $\pm$ 0.03  &  0.02 $\pm$ 0.03  &  $<$0.65              &  0.80    &  3.79  &   rgb-base  &   HARPS  \\
    NGC5822No102  &   5252 $\pm$  38  &  2.92 $\pm$ 0.22  &  1.37 $\pm$ 0.04  &  0.01 $\pm$ 0.03  &     1.45 $\pm$  0.13  &  1.57    &  5.40  &   rgb-base  &   UVES  \\
    NGC5822No224  &   5190 $\pm$  33  &  3.02 $\pm$ 0.09  &  1.38 $\pm$ 0.04  &  0.03 $\pm$ 0.03  &  $<$0.40              &  0.56    &  3.69  &   rgb-base  &   UVES  \\
    NGC5822No438  &   5157 $\pm$  33  &  3.10 $\pm$ 0.08  &  1.34 $\pm$ 0.04  &  0.05 $\pm$ 0.03  &  $<$0.40              &  0.56    &  2.95  &   rgb-base  &   UVES  \\
\hline
     NGC2423No20  &   5073 $\pm$  38  &  2.99 $\pm$ 0.10  &  1.38 $\pm$ 0.04  &  0.09 $\pm$ 0.03  &  $<$0.20              &  0.36    &  3.33  &      clump  &   UVES  \\
    NGC2423No240  &   5058 $\pm$  38  &  2.87 $\pm$ 0.11  &  1.45 $\pm$ 0.04  &  0.08 $\pm$ 0.03  &     0.67 $\pm$  0.16  &  0.83    &  3.13  &      clump  &   UVES  \\
      NGC2423No3  &   4592 $\pm$  72  &  2.33 $\pm$ 0.20  &  1.53 $\pm$ 0.08  & -0.03 $\pm$ 0.05  &     1.25 $\pm$  0.12  &  1.52    &  3.84  &       bump  &   UVES  \\
\hline
   IC4651No10393  &   4762 $\pm$  66  &  2.53 $\pm$ 0.15  &  1.36 $\pm$ 0.06  & -0.05 $\pm$ 0.05  & $<$-0.05              &  0.15    &  4.07  &      first  &   HARPS  \\
   IC4651No11453  &   5144 $\pm$  41  &  2.97 $\pm$ 0.08  &  1.62 $\pm$ 0.04  &  0.10 $\pm$ 0.03  &  $<$0.40              &  0.56    &  4.39  &    unknown  &   HARPS  \\
   IC4651No14527  &   4894 $\pm$  68  &  2.76 $\pm$ 0.14  &  1.43 $\pm$ 0.06  &  0.12 $\pm$ 0.05  &  $<$0.20              &  0.36    &  4.13  &      clump  &   HARPS  \\
    IC4651No7646  &   4965 $\pm$  45  &  2.74 $\pm$ 0.09  &  1.58 $\pm$ 0.05  &  0.06 $\pm$ 0.04  &  $<$0.30              &  0.46    &  3.75  &      clump  &   HARPS  \\
    IC4651No8540  &   4868 $\pm$  72  &  2.90 $\pm$ 0.14  &  1.48 $\pm$ 0.07  &  0.14 $\pm$ 0.05  &     0.67 $\pm$  0.18  &  0.83    &  3.54  &      clump  &   HARPS  \\
    IC4651No9025  &   4854 $\pm$  64  &  2.75 $\pm$ 0.13  &  1.46 $\pm$ 0.06  &  0.09 $\pm$ 0.05  &     0.70 $\pm$  0.18  &  0.86    &  3.56  &      clump  &   HARPS  \\
    IC4651No9122  &   4720 $\pm$  71  &  2.72 $\pm$ 0.15  &  1.40 $\pm$ 0.06  &  0.08 $\pm$ 0.04  & $<$-0.10              &  0.10    &  3.89  &      first  &   HARPS  \\
    IC4651No9791  &   4537 $\pm$  72  &  2.33 $\pm$ 0.19  &  1.48 $\pm$ 0.06  &  0.00 $\pm$ 0.04  &     1.14 $\pm$  0.13  &  1.41    &  4.40  &       bump  &   HARPS  \\
\hline
     NGC3680No13  &   4674 $\pm$  56  &  2.52 $\pm$ 0.14  &  1.41 $\pm$ 0.06  & -0.09 $\pm$ 0.04  &     1.17 $\pm$  0.12  &  1.44    &  2.67  &       bump  &   UVES  \\
     NGC3680No26  &   4668 $\pm$  47  &  2.48 $\pm$ 0.11  &  1.36 $\pm$ 0.05  & -0.08 $\pm$ 0.04  &     1.07 $\pm$  0.12  &  1.34    &  3.21  &       bump  &   UVES  \\
     NGC3680No41  &   4758 $\pm$  62  &  2.63 $\pm$ 0.14  &  1.41 $\pm$ 0.06  & -0.07 $\pm$ 0.04  &     0.69 $\pm$  0.18  &  0.89    &  3.02  &      first  &   UVES  \\
\hline
    NGC2682No164  &   4737 $\pm$  55  &  2.58 $\pm$ 0.14  &  1.46 $\pm$ 0.05  &  0.01 $\pm$ 0.04  &  $<$0.15              &  0.37    &  2.99  &      clump  &   UVES  \\
    NGC2682No266  &   4810 $\pm$  52  &  2.73 $\pm$ 0.12  &  1.52 $\pm$ 0.05  & -0.01 $\pm$ 0.03  &  $<$0.05              &  0.22    &  2.61  &      clump  &   UVES  \\
    NGC2682No286  &   4771 $\pm$  50  &  2.63 $\pm$ 0.13  &  1.49 $\pm$ 0.05  &  0.01 $\pm$ 0.03  &  $<$0.15              &  0.37    &  2.92  &      clump  &   UVES  \\
\hline
\noalign{\medskip} 
\end{tabular}
\end{table*}
\label{parameters}
\end{center}

\begin{center}
\begin{table*}
\caption{Masses, radii, ages, \logg, and luminosities from PARSEC isochrones for the stars in our sample.}
\centering
\begin{tabular}{lccccc}
\hline
\noalign{\medskip} 
Star & Mass & Radius & Age & \logg & Luminosity \\  
     & (M$_\odot$)  &  (R$_\odot$) & (Gyr) & (cm\,s$^{-2}$) & (log(L/L$_\odot$))  \\
\noalign{\medskip} 
\hline
\hline
    NGC3114No262  &   4.08 $\pm$ 0.09  &  30.10 $\pm$  3.61  &  0.168 $\pm$ 0.008  &   2.06 $\pm$ 0.10  &   2.74 $\pm$ 0.11  \\
      NGC3114No6  &   4.28 $\pm$ 0.15  &  40.87 $\pm$  3.72  &  0.157 $\pm$ 0.014  &   1.82 $\pm$ 0.07  &   2.97 $\pm$ 0.08  \\
     NGC3114No13  &   4.04 $\pm$ 0.05  &  26.25 $\pm$  0.75  &  0.172 $\pm$ 0.004  &   2.18 $\pm$ 0.02  &   2.63 $\pm$ 0.03  \\
    NGC3114No181  &   4.14 $\pm$ 0.10  &  38.24 $\pm$  3.30  &  0.165 $\pm$ 0.010  &   1.86 $\pm$ 0.07  &   2.85 $\pm$ 0.08  \\
    NGC3114No273  &   4.08 $\pm$ 0.01  &  24.69 $\pm$  0.55  &  0.174 $\pm$ 0.001  &   2.24 $\pm$ 0.01  &   2.67 $\pm$ 0.02  \\
\hline
    NGC4349No127  &   3.81 $\pm$ 0.23  &  36.98 $\pm$  4.89  &  0.203 $\pm$ 0.031  &   1.86 $\pm$ 0.10  &   2.81 $\pm$ 0.12  \\
    NGC4349No168  &   3.66 $\pm$ 0.12  &  20.12 $\pm$  1.28  &  0.224 $\pm$ 0.018  &   2.37 $\pm$ 0.05  &   2.47 $\pm$ 0.06  \\
    NGC4349No174  &   3.73 $\pm$ 0.16  &  29.80 $\pm$  3.66  &  0.217 $\pm$ 0.022  &   2.03 $\pm$ 0.09  &   2.67 $\pm$ 0.11  \\
    NGC4349No203  &   3.66 $\pm$ 0.11  &  19.88 $\pm$  1.15  &  0.226 $\pm$ 0.017  &   2.38 $\pm$ 0.04  &   2.48 $\pm$ 0.05  \\
      NGC4349No5  &   3.63 $\pm$ 0.11  &  21.13 $\pm$  2.68  &  0.225 $\pm$ 0.017  &   2.32 $\pm$ 0.10  &   2.50 $\pm$ 0.11  \\
     NGC4349No53  &   3.68 $\pm$ 0.10  &  26.83 $\pm$  2.31  &  0.228 $\pm$ 0.015  &   2.12 $\pm$ 0.06  &   2.64 $\pm$ 0.08  \\
      NGC4349No9  &   3.62 $\pm$ 0.11  &  20.08 $\pm$  1.81  &  0.227 $\pm$ 0.016  &   2.37 $\pm$ 0.07  &   2.51 $\pm$ 0.08  \\
\hline
     IC2714No110  &   3.01 $\pm$ 0.04  &  12.74 $\pm$  0.29  &  0.392 $\pm$ 0.011  &   2.68 $\pm$ 0.02  &   2.06 $\pm$ 0.03  \\
     IC2714No121  &   3.04 $\pm$ 0.08  &  17.05 $\pm$  1.21  &  0.379 $\pm$ 0.023  &   2.43 $\pm$ 0.06  &   2.21 $\pm$ 0.06  \\
     IC2714No126  &   3.02 $\pm$ 0.06  &  12.86 $\pm$  0.46  &  0.385 $\pm$ 0.019  &   2.68 $\pm$ 0.03  &   2.08 $\pm$ 0.04  \\
     IC2714No190  &   2.99 $\pm$ 0.05  &  13.78 $\pm$  1.23  &  0.391 $\pm$ 0.013  &   2.61 $\pm$ 0.08  &   2.11 $\pm$ 0.08  \\
     IC2714No220  &   3.03 $\pm$ 0.06  &  15.82 $\pm$  1.03  &  0.387 $\pm$ 0.017  &   2.49 $\pm$ 0.05  &   2.20 $\pm$ 0.06  \\
      IC2714No53  &   3.00 $\pm$ 0.04  &  12.67 $\pm$  0.22  &  0.393 $\pm$ 0.011  &   2.69 $\pm$ 0.01  &   2.08 $\pm$ 0.02  \\
       IC2714No5  &   3.00 $\pm$ 0.05  &  12.73 $\pm$  0.34  &  0.388 $\pm$ 0.016  &   2.69 $\pm$ 0.02  &   2.08 $\pm$ 0.03  \\
      IC2714No87  &   3.01 $\pm$ 0.04  &  12.69 $\pm$  0.24  &  0.393 $\pm$ 0.011  &   2.69 $\pm$ 0.01  &   2.11 $\pm$ 0.02  \\
\hline
    NGC2539No229  &   3.00 $\pm$ 0.06  &  12.47 $\pm$  0.55  &  0.399 $\pm$ 0.023  &   2.70 $\pm$ 0.03  &   2.04 $\pm$ 0.04  \\
    NGC2539No246  &   2.99 $\pm$ 0.06  &  12.45 $\pm$  0.54  &  0.399 $\pm$ 0.023  &   2.70 $\pm$ 0.03  &   2.05 $\pm$ 0.04  \\
    NGC2539No251  &   2.98 $\pm$ 0.06  &  12.41 $\pm$  0.51  &  0.400 $\pm$ 0.022  &   2.70 $\pm$ 0.03  &   2.04 $\pm$ 0.04  \\
    NGC2539No317  &   3.03 $\pm$ 0.09  &  12.86 $\pm$  0.87  &  0.383 $\pm$ 0.033  &   2.68 $\pm$ 0.05  &   2.07 $\pm$ 0.06  \\
    NGC2539No346  &   3.00 $\pm$ 0.09  &  12.87 $\pm$  0.84  &  0.386 $\pm$ 0.031  &   2.67 $\pm$ 0.04  &   2.06 $\pm$ 0.06  \\
    NGC2539No447  &   3.06 $\pm$ 0.12  &  13.49 $\pm$  1.14  &  0.372 $\pm$ 0.039  &   2.63 $\pm$ 0.06  &   2.08 $\pm$ 0.07  \\
    NGC2539No463  &   3.05 $\pm$ 0.12  &  13.78 $\pm$  1.30  &  0.375 $\pm$ 0.040  &   2.62 $\pm$ 0.07  &   2.10 $\pm$ 0.08  \\
    NGC2539No502  &   3.00 $\pm$ 0.07  &  12.58 $\pm$  0.66  &  0.393 $\pm$ 0.027  &   2.69 $\pm$ 0.04  &   2.05 $\pm$ 0.05  \\
    NGC2539No652  &   3.12 $\pm$ 0.11  &  17.74 $\pm$  1.73  &  0.372 $\pm$ 0.038  &   2.41 $\pm$ 0.07  &   2.26 $\pm$ 0.09  \\
\hline
    NGC6633No106  &   2.69 $\pm$ 0.01  &  10.01 $\pm$  0.08  &  0.525 $\pm$ 0.002  &   2.83 $\pm$ 0.01  &   1.86 $\pm$ 0.01  \\
    NGC6633No100  &   2.71 $\pm$ 0.04  &  10.58 $\pm$  0.36  &  0.512 $\pm$ 0.017  &   2.79 $\pm$ 0.02  &   1.88 $\pm$ 0.03  \\
    NGC6633No119  &   2.69 $\pm$ 0.02  &  10.00 $\pm$  0.01  &  0.525 $\pm$ 0.000  &   2.84 $\pm$ 0.01  &   1.87 $\pm$ 0.01  \\
    NGC6633No126  &   2.69 $\pm$ 0.00  &  10.00 $\pm$  0.02  &  0.525 $\pm$ 0.001  &   2.84 $\pm$ 0.01  &   1.87 $\pm$ 0.01  \\
\hline
      IC4756No44  &   2.63 $\pm$ 0.00  &   9.33 $\pm$  0.22  &  0.575 $\pm$ 0.002  &   2.89 $\pm$ 0.01  &   1.82 $\pm$ 0.03  \\
     IC4756No125  &   2.58 $\pm$ 0.03  &   9.43 $\pm$  0.21  &  0.573 $\pm$ 0.007  &   2.87 $\pm$ 0.01  &   1.81 $\pm$ 0.02  \\
      IC4756No38  &   2.58 $\pm$ 0.02  &   9.49 $\pm$  0.15  &  0.575 $\pm$ 0.002  &   2.88 $\pm$ 0.00  &   1.81 $\pm$ 0.02  \\
      IC4756No42  &   2.60 $\pm$ 0.03  &   9.15 $\pm$  0.11  &  0.575 $\pm$ 0.004  &   2.89 $\pm$ 0.01  &   1.80 $\pm$ 0.01  \\
\hline
    NGC2360No119  &   2.98 $\pm$ 0.25  &  12.59 $\pm$  2.26  &  0.393 $\pm$ 0.088  &   2.69 $\pm$ 0.12  &   2.05 $\pm$ 0.16  \\
     NGC2360No50  &   2.95 $\pm$ 0.24  &  12.36 $\pm$  2.19  &  0.402 $\pm$ 0.090  &   2.70 $\pm$ 0.12  &   2.04 $\pm$ 0.15  \\
     NGC2360No66  &   2.80 $\pm$ 0.19  &  10.89 $\pm$  1.66  &  0.468 $\pm$ 0.089  &   2.79 $\pm$ 0.10  &   1.96 $\pm$ 0.13  \\
     NGC2360No79  &   2.87 $\pm$ 0.22  &  11.54 $\pm$  1.90  &  0.438 $\pm$ 0.091  &   2.75 $\pm$ 0.11  &   1.98 $\pm$ 0.14  \\
      NGC2360No7  &   2.94 $\pm$ 0.23  &  12.18 $\pm$  2.13  &  0.408 $\pm$ 0.089  &   2.71 $\pm$ 0.12  &   2.03 $\pm$ 0.15  \\
     NGC2360No85  &   2.77 $\pm$ 0.19  &  10.68 $\pm$  1.54  &  0.482 $\pm$ 0.089  &   2.80 $\pm$ 0.09  &   1.92 $\pm$ 0.13  \\
     NGC2360No86  &   2.73 $\pm$ 0.21  &  12.98 $\pm$  1.56  &  0.514 $\pm$ 0.117  &   2.62 $\pm$ 0.08  &   2.05 $\pm$ 0.10  \\
     NGC2360No89  &   3.00 $\pm$ 0.25  &  12.89 $\pm$  2.36  &  0.383 $\pm$ 0.088  &   2.67 $\pm$ 0.12  &   2.07 $\pm$ 0.16  \\
\hline
    NGC5822No201  &   2.53 $\pm$ 0.11  &   8.68 $\pm$  0.83  &  0.628 $\pm$ 0.078  &   2.94 $\pm$ 0.06  &   1.78 $\pm$ 0.08  \\
      NGC5822No8  &   2.54 $\pm$ 0.12  &   9.07 $\pm$  0.84  &  0.615 $\pm$ 0.079  &   2.90 $\pm$ 0.06  &   1.76 $\pm$ 0.08  \\
    NGC5822No102  &   2.32 $\pm$ 0.07  &   7.34 $\pm$  0.47  &  0.787 $\pm$ 0.067  &   3.04 $\pm$ 0.04  &   1.61 $\pm$ 0.06  \\
    NGC5822No224  &   2.34 $\pm$ 0.08  &   7.51 $\pm$  0.52  &  0.774 $\pm$ 0.072  &   3.03 $\pm$ 0.05  &   1.61 $\pm$ 0.06  \\
    NGC5822No438  &   2.32 $\pm$ 0.07  &   7.40 $\pm$  0.44  &  0.798 $\pm$ 0.063  &   3.04 $\pm$ 0.04  &   1.59 $\pm$ 0.05  \\
\hline
     NGC2423No20  &   2.25 $\pm$ 0.03  &   7.08 $\pm$  0.24  &  0.894 $\pm$ 0.030  &   3.06 $\pm$ 0.03  &   1.53 $\pm$ 0.03  \\
    NGC2423No240  &   2.28 $\pm$ 0.06  &   7.45 $\pm$  0.40  &  0.853 $\pm$ 0.057  &   3.02 $\pm$ 0.04  &   1.57 $\pm$ 0.05  \\
      NGC2423No3  &   2.26 $\pm$ 0.07  &  13.01 $\pm$  1.11  &  0.850 $\pm$ 0.068  &   2.54 $\pm$ 0.07  &   1.93 $\pm$ 0.08  \\
\hline
   IC4651No10393  &   2.06 $\pm$ 0.11  &   9.62 $\pm$  1.08  &  1.185 $\pm$ 0.173  &   2.76 $\pm$ 0.09  &   1.71 $\pm$ 0.10  \\
   IC4651No11453  &   2.65 $\pm$ 0.12  &   9.60 $\pm$  0.88  &  0.558 $\pm$ 0.068  &   2.87 $\pm$ 0.06  &   1.81 $\pm$ 0.08  \\
   IC4651No14527  &   2.09 $\pm$ 0.09  &   8.81 $\pm$  0.51  &  1.191 $\pm$ 0.154  &   2.84 $\pm$ 0.04  &   1.67 $\pm$ 0.05  \\
    IC4651No7646  &   2.31 $\pm$ 0.15  &  10.25 $\pm$  0.85  &  0.898 $\pm$ 0.187  &   2.75 $\pm$ 0.05  &   1.82 $\pm$ 0.07  \\
    IC4651No8540  &   2.09 $\pm$ 0.09  &   8.87 $\pm$  0.53  &  1.193 $\pm$ 0.154  &   2.84 $\pm$ 0.04  &   1.67 $\pm$ 0.06  \\
    IC4651No9025  &   2.09 $\pm$ 0.10  &   8.87 $\pm$  0.57  &  1.185 $\pm$ 0.160  &   2.84 $\pm$ 0.04  &   1.67 $\pm$ 0.06  \\
    IC4651No9122  &   2.06 $\pm$ 0.09  &   8.90 $\pm$  0.68  &  1.223 $\pm$ 0.150  &   2.83 $\pm$ 0.06  &   1.64 $\pm$ 0.07  \\
    IC4651No9791  &   2.00 $\pm$ 0.09  &  13.70 $\pm$  1.38  &  1.229 $\pm$ 0.158  &   2.44 $\pm$ 0.09  &   1.96 $\pm$ 0.09  \\
\hline
     NGC3680No13  &   1.90 $\pm$ 0.07  &  11.61 $\pm$  1.03  &  1.345 $\pm$ 0.133  &   2.58 $\pm$ 0.08  &   1.85 $\pm$ 0.08  \\
     NGC3680No26  &   1.89 $\pm$ 0.05  &  11.33 $\pm$  0.92  &  1.372 $\pm$ 0.109  &   2.58 $\pm$ 0.07  &   1.83 $\pm$ 0.07  \\
     NGC3680No41  &   1.96 $\pm$ 0.09  &   9.30 $\pm$  1.12  &  1.296 $\pm$ 0.152  &   2.77 $\pm$ 0.10  &   1.68 $\pm$ 0.11  \\
\hline
    NGC2682No164  &   1.62 $\pm$ 0.14  &  10.67 $\pm$  0.55  &  2.404 $\pm$ 0.585  &   2.57 $\pm$ 0.07  &   1.80 $\pm$ 0.05  \\
    NGC2682No266  &   1.74 $\pm$ 0.17  &  10.30 $\pm$  0.65  &  1.952 $\pm$ 0.520  &   2.63 $\pm$ 0.08  &   1.79 $\pm$ 0.06  \\
    NGC2682No286  &   1.69 $\pm$ 0.17  &  10.69 $\pm$  0.59  &  2.160 $\pm$ 0.596  &   2.58 $\pm$ 0.07  &   1.81 $\pm$ 0.05  \\
\hline
\noalign{\medskip} 
\end{tabular}
\end{table*}
\label{padova}
\end{center}

\clearpage

\section{Stellar spectra for clusters with enhanced Li}  \label{app:plots}

\begin{figure}[h]
\centering
\includegraphics[width=9.0cm]{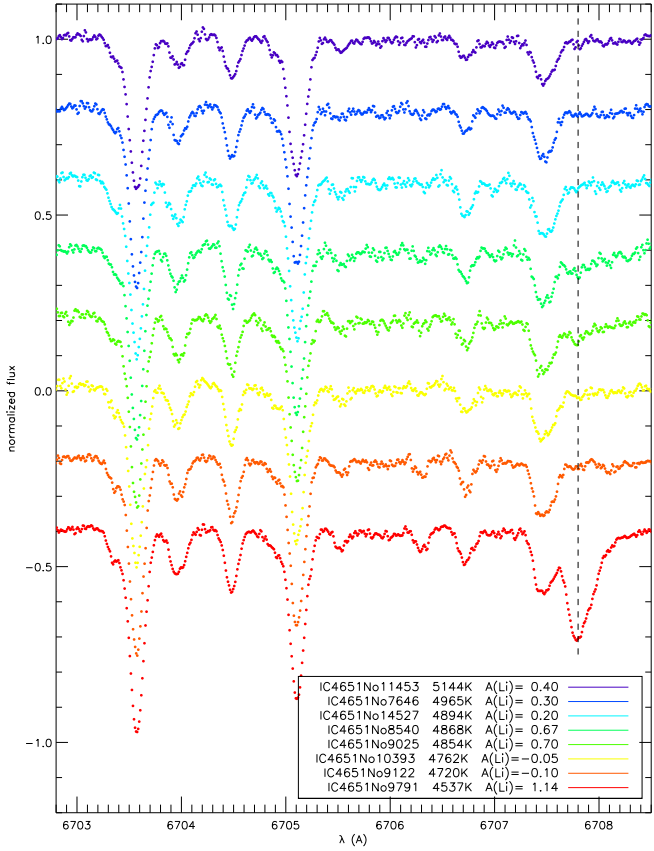}
\caption{Spectral region around Li in the open cluster IC4651. The position of the Li line is shown with a dashed line.} 
\label{IC4651}
\end{figure}

\begin{figure}
\centering
\includegraphics[width=9.0cm]{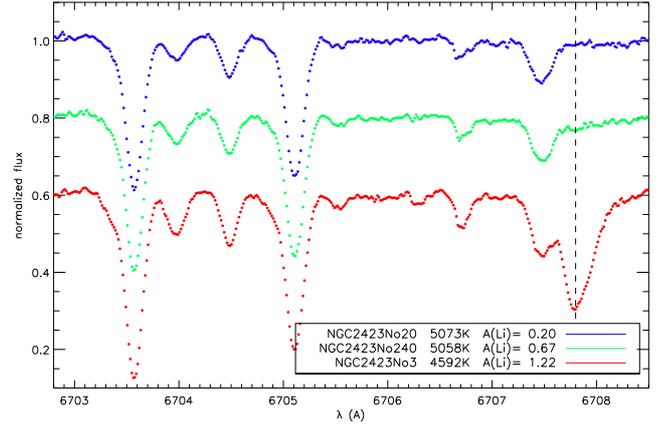}
\includegraphics[width=7.0cm,angle=270]{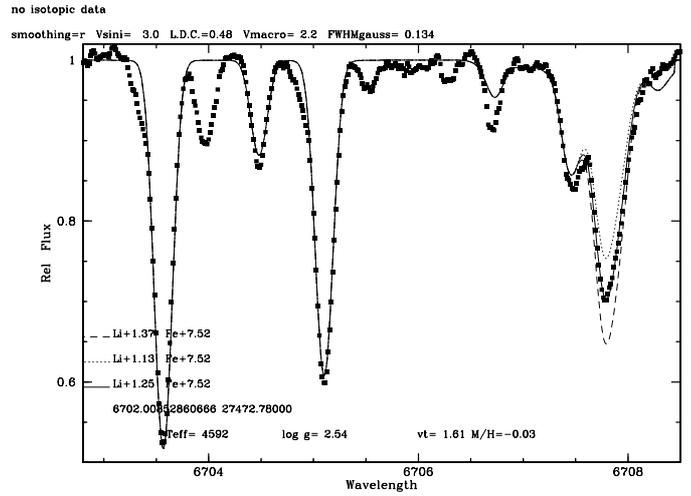}
\caption{Spectral region around Li in the open cluster NGC2423 and spectral synthesis around the Li line for NGC2423No3.} 
\label{NGC2423}
\end{figure}

\begin{figure}
\centering
\includegraphics[width=9.0cm]{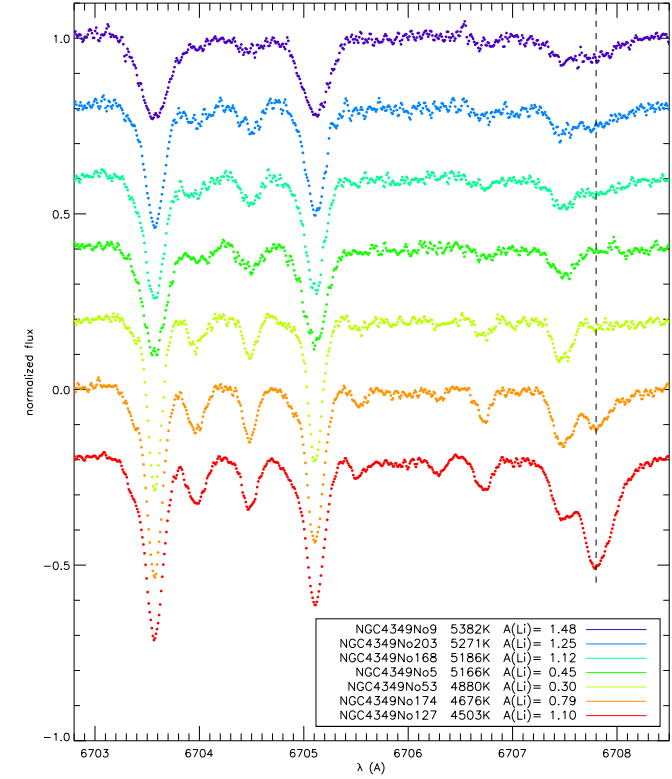}
\includegraphics[width=7.0cm,angle=270]{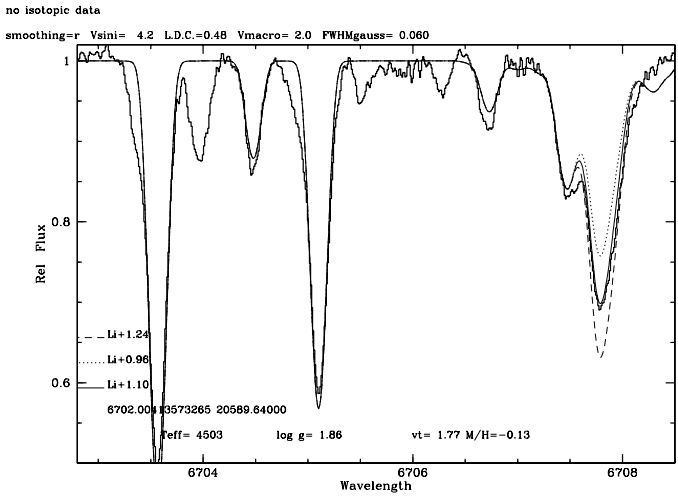}
\caption{Spectral region around Li in the open cluster NGC4349 and spectral synthesis around the Li line for NGC4349No127.} 
\label{NGC4349}
\end{figure}

\begin{figure}
\centering
\includegraphics[width=9.0cm]{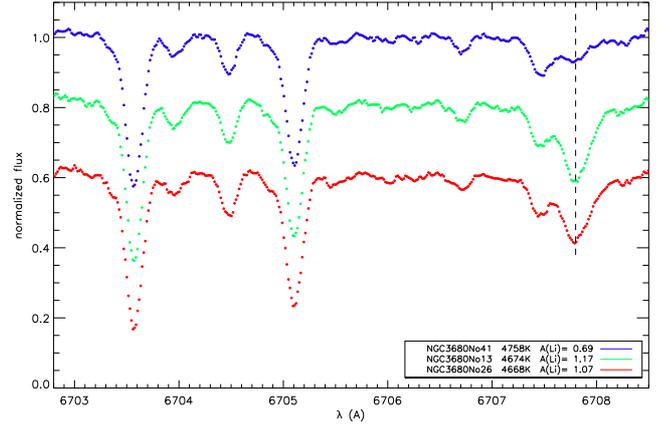}
\includegraphics[width=7.0cm,angle=270]{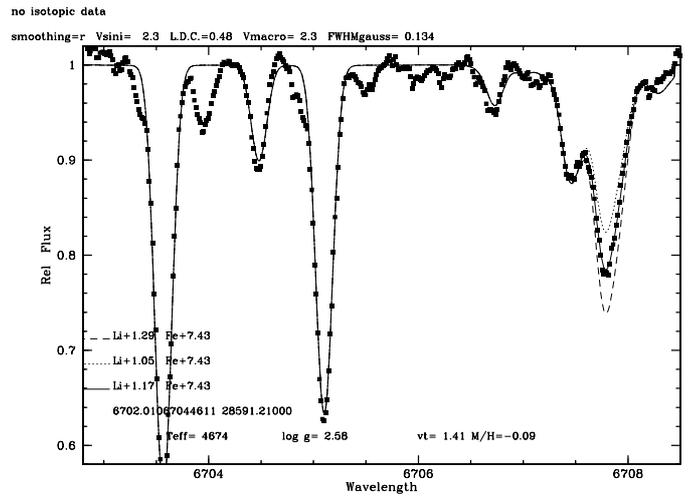}
\includegraphics[width=7.0cm,angle=270]{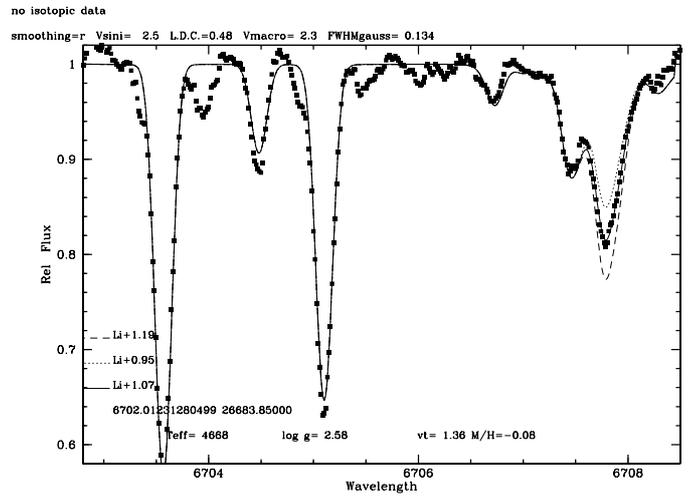}\caption{Spectral region around Li in the open cluster NGC3680 and spectral synthesis around the Li line for NGC3680No13 (middle panel) and NGC3680No26 (bottom panel).} 
\label{NGC3680}
\end{figure}

\clearpage

\end{document}